\documentclass[a4paper,11pt]{article}
\pdfoutput=1 
\usepackage{physics}

\usepackage{jheppub} 
\usepackage[T1]{fontenc} 
\usepackage{slashed}
\newcommand{\be}{\begin{equation}}
\newcommand{\ee}{\end{equation}}
\usepackage{tikz}
\usetikzlibrary{shapes.geometric, arrows}
\usepackage{tikz-cd}
\tikzstyle{process} = [rectangle, minimum width=3cm, minimum height=1cm, text centered, text width=3cm, draw=black, fill=orange!30]
\tikzstyle{arrow} = [thick,->,>=stealth]

\usepackage{ucs}
\usepackage{hyperref}
\usepackage[utf8x]{inputenc}
\usepackage{amsmath,bm}
\usepackage{amsfonts}
\usepackage{comment}
\usepackage{amssymb}
\usepackage{array}
\usepackage{epsfig}
\usepackage{slashed}
\usepackage{lipsum}
\usepackage{hyperref}
\hypersetup{
	colorlinks=true,
	linkcolor=blue,
	filecolor=red,
	urlcolor=blue,
	citecolor=red
}

\title{4D Flat-space scattering amplitude /$CFT_3$ correlator correspondence revisited}
\author{Sachin Jain,}
\author{Abhishek Mehta}

\affiliation{Indian Institute of Science Education and Research, Homi Bhabha Road, Pashan, Pune 411 008, India}

\emailAdd{sachin.jain@iiserpune.ac.in}
\emailAdd{abhishek.mehta@students.iiserpune.ac.in}

\abstract{In this paper, we make connection between CFT$_3$ three point correlation function of conserved currents and 4D three point amplitude of general spin massless gauge field explicit. We do so by taking flat space limit of momentum space CFT correlation function and show how they reproduce flat space amplitudes. We then point out a mismatch between  number of independent structures in 3D CFT correlator of conserved currents and 4D flat space covariant vertex of massless higher spin fields.  This is in contrast with  general expectation that counting of 3d CFT correlator and 4d flat space amplitude should match. This mismatch is even more pronounced in spinor helicity variables. 
	
We also point out an interesting relation between parity even and parity odd flat space amplitude in momentum space. This observation helps  us to construct a new momentum space CFT strtucture which  accounts for the mismatch.  However we should mention that 
this extra CFT structure can't be constructed out of conserved currents and hence counting  mismatch between CFT correlation of conserved currents and flat space amplitude of massless gauge field persists. Story in spinor helicity variable is more complicated and is discussed in detail.  We further comment on the connection of CFT correlation function in spinor helicity variables to AdS amplitudes in spinor helicity variables and light cone variables.

}

\begin{document}
	
	\maketitle
	
	\raggedbottom
	
	\section{Introduction}
	Conformal field theory (CFT) plays an important role in physics. One of the main quantities of interest in CFT is its correlation function of operators. In the last decade, CFT has seen tremendous progress in position space, however very little progress has been made in momentum space. Even though relatively less explored,
exploration of CFT correlation function in momentum space \cite{Coriano:2013jba,Bzowski:2013sza,Bzowski:2015pba,Bzowski:2017poo,Bzowski:2018fql,Gillioz:2019lgs,Isono:2019ihz} in three dimension \cite{Maldacena:2011nz,Baumann:2019oyu,Baumann:2020dch,Jain:2020rmw,Jain:2020puw,Jain:2021wyn,Jain:2021vrv,Caron-Huot:2021kjy,Jain:2021gwa,Gandhi:2021gwn} has already led to new understanding of structures of correlation function which were not understood in position space \cite{Farrow:2018yni,Jain:2021wyn,Caron-Huot:2021kjy,Jain:2021gwa}.
Momentum space correlation function also find application in cosmology \cite{Maldacena:2011nz,Baumann:2019oyu,Mata:2012bx,Kundu:2014gxa,Sleight:2019hfp,Pajer:2020wxk}.
Another remarkable feature of the momentum space CFT correlation function is its connection to flat-space amplitude \cite{Penedones:2010ue,Raju:2012zr}. Any $d$-dimensional three-point CFT correlator can be shown to give rise to $d+1$-dimensional three-point flat-space amplitude in the flat-space limit. The connection between momentum space helicity structure and three-point coupling in higher spin AdS theory was also explored in \cite{Metsaev:2018xip,Nagaraj:2018nxq,Nagaraj:2019zmk}. The relation between CFT correlator and AdS amplitude were discussed in \cite{Skvortsov:2018uru}.

Any three-point CFT correlator of conserved current has a maximum of three structures, two parity-even and one parity-odd structure \cite{Giombi:2011rz,Maldacena:2012sf}. Two parity-even structures can be obtained from free bosonic or free fermionic theory whereas the parity-odd structures can also be obtained from local Lagrangian such as Chern-Simons matter theories \cite{Giombi:2011kc,Aharony:2011jz}. On the other hand, there are two parity-even amplitude structures and two parity-odd covariant amplitude structures \cite{Bengtsson:1986kh,Conde:2016izb}. Interestingly, it is known that in spinor helicity variables there are four parity even and four parity odd amplitudes for massless spinning fields \cite{PhysRevLett.56.2459,Elvang:2013cua}. In AdS as well there are four parity even and four parity odd amplitudes in both spinor helicity variables and in light cone gauge \cite{Metsaev:2018xip,Nagaraj:2018nxq,Nagaraj:2019zmk}. This is summarised in the following table.
\newpage
\begin{table}[h!]
	\begin{tabular}{||c | c | c | c | c ||}
		\hline
		\multicolumn{5}{|c|}{Amplitude of massless higher spin fields in 4D vs CFT correlator of conserved currents in 3D}\\
		\hline\hline
		$\mathbf{s_1 \leq s_2 \leq s_3}$& \begin{tabular}{@{}c@{}}\bf{Covariant}\\\bf{Cubic vertex} \end{tabular} & \begin{tabular}{@{}c@{}}\bf{Spinor Helicity}\\\bf{Amplitude} \end{tabular} & \begin{tabular}{@{}c@{}}\bf{CFT Correlator}\\\bf{of conserved currents} \end{tabular} & \begin{tabular}{@{}c@{}}\bf{AdS amplitude}\\\bf{in spinor helicity }\\\bf{ and light cone} \end{tabular}\\ [0.5ex]
		\hline
		 $s_3\le s_1+s_2$ & 2 even, 2 odd & 4 even , 4 odd & 2 even, 1 odd & 4 even , 4 odd\\
		\hline
		 $s_ 3> s_1+s_2$ & 2 even, 2 odd & 4 even , 4 odd & 2 even & 4 even , 4 odd \\
		\hline
	\end{tabular}
	\caption{Comparison of flat space/AdS amplitude and CFT correlator. Special cases involving one or more scalar and two or three equal spins are discussed in the main text.}
	\label{intamp}
\end{table}

The table immediately makes it clear that there is gross mismatch in number of independent structures of amplitude in flat space, AdS and CFT correlator. The mismatch between flat space covariant vertex and spinor helicity amplitude in 4D was already pointed out in \cite{Bengtsson:2014qza,Conde:2016izb}. Amplitudes that cannot be written as local Lorentz covariant vertices
were written in \cite{Benincasa:2011pg} in flat space. Interestingly, this puzzle was reconsidered in light-cone gauge in $AdS_4$ in \cite{Metsaev:2018xip} where it was shown that in the light cone gauge one can construct as many amplitudes as in spinor helicity variables.

As can be seen in the table \ref{intamp}, the number of CFT correlators is less than the number of flat space covariant amplitudes or the flat space spinor helicity amplitude. This immediately raises the question on validity of the flat space amplitude/ CFT correlator correspondence. 
Let us concentrate on the case of\footnote{For out side the triangle $s_3> s_1+s_2,$ see table \ref{intamp} for counting. In this case, if we consider slightly broken HS currents, then one gets one parity odd structure, which is still a mismatch with flat space amplitude.} $s_3\le s_1+s_2.$ In this case as is well known, see \cite{Giombi:2011rz,Costa:2011mg}, three are total three structures for CFT correlators of conserved currents two parity even and one parity odd \footnote{  
However if we allow for non conserved currents then there are more allowed structures \cite{Costa:2011mg,Kravchuk:2016qvl,Hebbar:2020ukp}. The point of interest of this paper is three point function of conserved currents for which how the counting works for flat space amplitude and CFT correaltors is not discussed in these papers. }.  This implies, in this case we see there is one less parity odd structure in CFT as compared to covariant vertex. Interestingly, we point out a CFT structure that in the flat space limit reproduces the missing flat space amplitude.
 To do this we first show that parity odd flat space covariant amplitude can be constructed starting from parity even flat space covariant amplitude by what we call as epsilon transformation. We show that by using same epsilon transform on some combination of parity even free fermion and free boson CFT correlator of conserved currents, we can construct a new parity odd CFT correlator which in the flat space limit reproduces the correct flat space covariant parity odd vertex. Even though, this new parity odd CFT correlator is constructed out of epsilon transform of correlation of conserved current, the resulting correlator is not that of conserved currents, as will be discussed. The mismatch in number of CFT structure and flat space spinor helicity amplitude is even larger. 
 In this case as well,  we could construct CFT correlation function which in the flat space limit does not give rise to covariant vertex but gives correct flat space spinor helicity amplitude. These extra CFT correlators have some unusual properties that they are not consistent with position space OPE limit.

The paper is organized as follows. In Section \ref{rvw1}, we give a brief review of various flat-space vertices in covariant notation. In Section \ref{eptr}, we introduce the epsilon transformation that allows us to go from parity-even to parity-odd structures for covariant vertex while preserving gauge-invariance. In section \ref{spampst} we discuss non-perturbative spinor helicity amplitudes and discuss the mismatch with covariant vertex.
In section \ref{fltsplmt}, we show how the CFT correlator maps to the various flat-space covariant vertices. In section \ref{nptod}, we make use of the epsilon transformation to propose a new CFT structure that in the flat-space limit gives the extra parity-odd amplitude and discuss several examples. In subsection section \ref{rpuza}, we discuss the mismatch between the flat-space limit of CFT correlator in spinor helicity variable and four-dimensional flat space amplitude in spinor helicity variables and discuss how CFT correlators can be constructed which in the flat space limit does not give rise to covariant vertex but gives correct flat space spinor helicity amplitude. In section \ref{adsfltab}, after reviewing some known results in amplitude in $AdS_4$ and their connection to CFT correlator, we discuss connection to CFT correlator in the light of results presented in this paper.
In section \ref{disc} we discuss the results obtained in this paper, some future directions, and double copy relations. In Appendix \ref{ampflt}, we discuss some explicit examples of flat-space amplitudes. In Appendix \ref{CFTfta}, we discuss various CFT correlation functions and their flat-space limits. In Appendix \ref{fltsp1} we discuss the flat space limit of CFT correlators in spinor helicity variables. Appendix \ref{npodapd} is devoted to understanding more on parity odd CFT correlators that we constructed by doing epsilon transformation. Appendix \ref{411ab} is devoted to understanding the missing CFT structures in spinor helicity variables for the example of $\langle JJ J_4\rangle$. In Appendix \ref{idnty}, we discuss various identities which are useful in the main text. The last Appendix \ref{zeroamp} discusses the special case of flat space amplitude and CFT correlator in spinor helicity variables with net helicity zero configuration.
\section{Brief review of flat-space covariant vertex }\label{rvw1}
In this section, we review known results for flat-space amplitude in four dimensions for massless particles. We follow closely the notation and discussion in \cite{Conde:2016izb}.

\subsection{Covariant vertex}
Consider a generic action of the form
\begin{align}
S^{(3)} = \int d^4x C(\partial_{x_I}, \partial_{u_I})\phi^1(x_1, u_1)\phi^2(x_2, u_2)\phi^3(x_3, u_3)\bigg|_{u_I = 0, x_I = x}
\end{align}
where
\begin{align}
\phi(x, u) = \phi_{\mu_1\cdots\mu_s}(x)u^{\mu_1}\cdots u^{\mu_s}
\end{align}
are higher spin fields and $u_I$ are some some auxiliary variables and $C$ is some operator which makes the higher-spin fields contract with each other with various derivatives forming a cubic interaction. Under higher spin symmetry and gauge invariance, $C$ can be made up of only certain parity-even and parity-odd structures namely
\begin{table}[h!]
\begin{tabular}{||c | c | c ||}
\hline
\multicolumn{3}{|c|}{Flat-space amplitudes Building blocks}\\
\hline\hline
\begin{tabular}{@{}c@{}}\bf{Parity} \end{tabular} & \begin{tabular}{@{}c@{}}\bf{Position space} \end{tabular} & \bf{Momentum space}\\ [0.5ex]
\hline
\begin{tabular}{@{}c@{}} Even \end{tabular}&\begin{tabular}{@{}c@{}}\bf{$Y_I = \partial_{u_I}.\partial_{x_{I+1}}$} \\ $Z_I = \partial_{u_{I+1}}.\partial_{u_{I-1}}$\end{tabular} &\begin{tabular}{@{}c@{}}\bf{$Y_I = z_I.p_{I+1}$} \\ $Z_I = z_{I+1}.z_{I-1} $\end{tabular} \\
\hline
\begin{tabular}{@{}c@{}}Odd \end{tabular}&\begin{tabular}{@{}c@{}}$V_I = \epsilon^{\mu\nu\rho\sigma}\partial_{u^{\mu}_{I+1}}\partial_{x^{\nu}_{I+1}}\partial_{u^{\rho}_{I-1}}\partial_{x^{\sigma}_{I-1}}$ \\ $W_I = \epsilon^{\mu\nu\rho\sigma}\partial_{u^{\mu}_{1}}\partial_{u^{\nu}_{2}}\partial_{u^{\rho}_{3}}\partial_{x^{\sigma}_{I}}$\end{tabular} &\begin{tabular}{@{}c@{}}\bf{ $V_I = \epsilon(z_{I+1}p_{I+1}z_{I-1}p_{I-1})$ } \\ $W_I = \epsilon(z_1z_2z_3p_I)$\end{tabular} \\ [1ex]
\hline
\end{tabular}
\caption{Building blocks for flat-space amplitude in 4d. $\epsilon$ is 4D totally antisymmetric tensor and the notation $\epsilon(xyzw)=\epsilon_{\mu\nu\rho\sigma}x^{\mu}y^{\nu}z^{\rho} w^{\sigma}.$}
\label{Tamp1}
\end{table}
\newpage
The gauge symmetry and higher spin symmetry, constrains the possible amplitudes to be

\begin{table}[h!]
\begin{tabular}{||c | c | c ||}
\hline
\multicolumn{3}{|c|}{Flat-space amplitudes}\\
\hline\hline
\begin{tabular}{@{}c@{}}\bf{Parity} \end{tabular} & \begin{tabular}{@{}c@{}}\bf{Minimal } \end{tabular} & \bf{Non-Minimal}\\ [0.5ex]
\hline
\begin{tabular}{@{}c@{}} parity-even \end{tabular}& $g_{m, e}G^{s_1}Y^{s_2-s_1}_2Y^{s_3-s_1}_3$ & $g_{nm, e}Y^{s_1}_1Y^{s_2}_2Y^{s_3}_3$ \\
\hline
\begin{tabular}{@{}c@{}}parity-odd \end{tabular}&$ g_{m, o} V_1 G^{s_1}_1Y^{s_2-s_1-1}_2Y^{s_3-s_1-1}_3$ & $g_{nm, o}V_1Y^{s_1}_1Y^{s_2-1}_2Y^{s_3-1}_3$ \\ [1ex]
\hline
\end{tabular}
\caption{Amplitudes in 4d. The subscript "m" means minimal, "nm" means non-minimal, "e" means parity even, "o" means parity odd.}
\label{Tamp2}
\end{table}
with $G= Y_1Z_1+Y_2Z_2 + Y_3 Z_3.$ We have also assumed that $s_3 > s_2, s_1$ with out loss of generality. Minimal and non-minimal distinction comes from number of derivatives or number of momentum factors present in the amplitudes.

Notice that for the odd case there is an issue when spins are coincident i.e. $s_i = s_j$, one obtains negative powers of derivatives which is not possible due to locality. For the case of $s_1 = s_2 < s_3$, one may use the identity
\begin{align}
V_{1} G^{s_{1}} Y_{2}^{-1} Y_{3}^{s_{3}-s_{1}-1} \approx \frac{1}{2}\left[V_{1} Z_{2}-V_{2} Z_{1}+\left(W_{2}-W_{1}\right) Y_{3}\right] G^{s_{1}-1} Y_{3}^{s_{3}-s_{1}-1}\label{oddamp}
\end{align}
to remove the negative powers. Such identities are derived using Schouten identities and momentum conservation. For the case of coincidence of $s_1 = s_2 < s_3$, notice that (\ref{oddamp}) is anti-symmetric under the $1\leftrightarrow 2$ exchange. This implies for parity odd minimal amplitude to be non-zero for this case, we do require Chan-Paton factor which is antisymmetric in exchange of $1 \leftrightarrow 2$ indices. Together with Chan-Paton factor and \eqref{oddamp}, the amplitude becomes symmetric.

For the case $s_1 = s_2 = s_3$, it can be shown that the negative powers for parity odd minimal amplitude cannot be removed \cite{Conde:2016izb}. In this case the cubic vertices with negative powers are simply dropped. The final results for special cases are summarized below in the table.
\begin{table}[h!]
\begin{tabular}{||c | c | c ||}
\hline
\multicolumn{3}{|c|}{parity-odd flat-space amplitudes: special cases}\\
\hline\hline
\begin{tabular}{@{}c@{}}\bf{Configuration} \end{tabular} & \begin{tabular}{@{}c@{}}\bf{Minimal } \end{tabular} & \bf{Non-Minimal}\\ [0.5ex]
\hline
\begin{tabular}{@{}c@{}}$s_1=s_2<s_3$ \end{tabular}& $g_{m, o} \left(V_{1} Z_{2}-V_{2} Z_{1}+\left(W_{2}-W_{1}\right) Y_{3}\right) G^{s_{1}-1} Y_{3}^{s_{3}-s_{1}-1}$ & $g_{nm, o}V_1 Y^{s_1}_1Y^{s_1-1}_2Y^{s_3-1}_3$ \\
\hline
\begin{tabular}{@{}c@{}} $s_1=s_2=s_3$ \end{tabular}& $\cross$ & $g_{nm, o}V_1Y^{s_1}_1Y^{s_1-1}_2Y^{s_1-1}_3$ \\ [1ex]
\hline
\end{tabular}
\caption{parity-odd special cases of amplitudes in 4d. The parity-even part of the amplitude is as given in Table \ref{Tamp2}.}
\label{Tamp2a}
\end{table}
\\
\\
Let us note that, the discussion involving scalar is simpler and can be obtained from results in Table \ref{Tamp2}.
Some explicit examples are worked in detail in the appendix \ref{ampflt}.

In the next section we show an interesting relation between parity even and parity odd part of the covariant vertex discussed till now.

\subsection{Epsilon transformation}\label{eptr}
In this section, we introduce what we call epsilon transform which maps parity-even amplitude to parity-odd amplitude and vice-versa.
In 4D one may work with the following choice of momenta and polarization
\begin{align}\label{3d2d}
p^{\mu} = (k,k^i) \quad z^{\mu} = (0, z^i)
\end{align}
with this choice the momentum space expressions in Table \ref{Tamp1} takes the form
\begin{align}
& V_I =\epsilon(z_{I+1}z_{I-1}k_{I-1})k_{I+1}-\epsilon(z_{I+1}z_{I-1}k_{I+1})k_{I-1}, \quad Y_I = z_{I}.k_{I+1}\nonumber\\
&Z_{I} = z_{I+1}.z_{I-1}, \quad p_I.p_J = -k_I k_J + k_I.k_J = 0\label{mandv}
\end{align}
where we have used three dimensional epsilon in the last expression, more precisely $\epsilon^{0\mu\nu\rho}=\epsilon^{\mu\nu\rho}$ where index $0$ is time direction.
Let us define an epsilon transformation \cite{Jain:2021gwa} which is given by
\begin{align}
z^i \to \frac{\epsilon^{z k i}}{k}
\end{align}
The $\epsilon$-transformation can also be implemented by a differential operator
\begin{align}\label{diffop}
[O_{\epsilon}]_I = \frac{1}{k_I}\epsilon(z_I k_I \frac{\partial}{\partial z_I})
\end{align}
which acts on parity-even gauge-invariant structures to give parity-odd gauge invariant structures i.e
\begin{align}
O_{\epsilon}: \mathcal{M}_{m, e} &\to \mathcal{M}_{m, o}\nonumber\\
O_{\epsilon}: \mathcal{M}_{nm, e} &\to \mathcal{M}_{nm, o}.
\end{align}
We show this explicitly below. Consider the epsilon transformation of the even minimal amplitude
\begin{align}
& [O_{\epsilon}]_2\mathcal{M}_{m, e} = s_1G^{s_1-1}Y^{s_2-s_1}_2Y^{s_3-s_1}_3[O_{\epsilon}]_2G+(s_2-s_1) G^{s_1}Y^{s_2-s_1-1}_2Y^{s_3-s_1}_3[O_{\epsilon}]_2Y_2 \label{S1}
\end{align}
Now, it can be shown \footnote{Refer to the Appendix C for its derivation.} that
\begin{align}
&Y_2Y_3 [O_{\epsilon}]_2G = -GV_1 \label{SID2}\\
&Y_3[O_{\epsilon}]_2Y_2 = -V_1\label{SID3}
\end{align}
which when used in (\ref{S1}) gives
\begin{align}\label{sid4a}
[O_{\epsilon}]_2\mathcal{M}_{m, e} = -s_2 V_1G^{s_1}Y^{s_2-s_1-1}_2Y^{s_3-s_1-1}_3=\mathcal{M}_{m, o}
\end{align}
which is precisely the odd minimal amplitude. Similarly, for the even non-minimal amplitude we have
\begin{align}
[O_{\epsilon}]_2\mathcal{M}_{nm, e} = s_2 Y^{s_1}_1 Y^{s_2-1}_2 Y^{s_3}_3[O_{\epsilon}]_2Y_2 = -s_2V_1 Y^{s_1}_1 Y^{s_2-1}_2 Y^{s_3-1}_3=\mathcal{M}_{nm, o}
\end{align}
where in the second equality (\ref{SID3}) was used and we have precisely obtained the odd non-minimal amplitude given in \ref{Tamp2}. Let us now consider some special examples to illustrate this.

\section*{All equal spin: $s_1=s_2=s_3$}
Notice when $s_1 = s_2 = s_3 = s$ for minimal even amplitude, from (\ref{sid4a}) we get
\begin{align}
[O_{\epsilon}]_2\mathcal{M}^{sss}_{m, e} = -s G^{s-1}Y^{-1}_2Y^{-1}_3
\end{align}
These negative powers cannot be removed by any Schouten identities or degeneracies, therefore, we conclude that parity odd minimal gauge-invariant vertex does not exist. This is consistent with table \ref{Tamp2a}.

For the case of $s_1 = s_2 = s_3 = 2$, from (\ref{cubv}) we have
\begin{align}\label{222cv1}
&\mathcal{M}^{222}_{m} = g_{m, e} G^2 \nonumber\\
&\mathcal{M}^{222}_{nm} = g_{nm, e} Y^2_1 Y^2_2 Y^2_3+ g_{nm, o} V_1 Y^2_1 Y_2 Y_3
\end{align}
It is easy to show that non-minimal parity even and parity odd terms are related by epsilon transform.
However, if we look at the epsilon transform of the minimal even amplitude we get
\begin{align}
[\mathcal{M}']^{222}=([O_{\epsilon}]_1+[O_{\epsilon}]_2+[O_{\epsilon}]_3)G^2 = -\frac{G}{k_1k_2k_3}(k_2k_3 \epsilon(z_1k_1k_2)z_2.z_3+\text{cyclic terms}) \label{epsliontmin}
\end{align}
which cannot be re-written as a $4D$ Lorentz invariant structure. Hence it is a not a valid covarint amplitude. This is consistent with the fact that there can not exist any parity odd minimal covariant vertex for an equal spin case.

\section*{Two equal spin: $s_1=s_2\ne s_3$}
Another coincidence point of concern is $s_1 = s_2 = s$, where we obtain
\begin{align}
[O_{\epsilon}]_2\mathcal{M}^{sss_3}_{ m, e} = -s V_1 G Y^{-1}_2 Y^{s_3-s-1}_3
\end{align}
These negative powers can be removed by using the Schouten identities \cite{Jain:2021wyn} and re-write them as
\begin{align}\label{sss3}
[O_{\epsilon}]_2\mathcal{M}^{sss_3}_{m, e} = -s\frac{1}{2}\left[V_{1} Z_{2}-V_{2} Z_{1}+\left(W_{2}-W_{1}\right) Y_{3}\right] G^{s-1} Y_{3}^{s_{3}-s-1}
\end{align}
Due to the nature of the Schouten identities in 4D momentum space, an anti-symmetrization in epsilon transforms at $1, 2$ was not necessary to derive the above, but one can in principle still use anti-symmetrization and obtain the same result using (\ref{SID6})
\begin{align}
( [O_{\epsilon}]_1-[O_{\epsilon}]_2)\mathcal{M}^{sss_3}_{m, e} = s\left[V_{1} Z_{2}-V_{2} Z_{1}+\left(W_{2}-W_{1}\right) Y_{3}\right] G^{s-1} Y_{3}^{s_{3}-s-1}
\end{align}
However, as we will see later, this anti-symmetrization of the epsilon transforms is necessary at the level of the CFT correlator because not all Schouten identities in 4D momentum space carry on to the 3D momentum space. The epsilon transform can be done at any operator in the correlator, however, for the case of $s_1 = s_2 $, some care needs to be taken. Notice that (\ref{sss3}) is anti-symmetric under $1\leftrightarrow 2$ while its minimal even counterpart is symmetric under the exchange. This is due to the presence of Chan-Paton factors for the case of $s_1 = s_2$. Therefore, an epsilon transform of minimal even for $s_3$ will give zero. Hence, when Chan-Paton factors are involved, the epsilon transform of $s_3$ must be avoided.

\section{Amplitude in spinor helicity variables}\label{spampst}
Let us now review expressions of amplitude written directly in spinor helicity variables.
The most general cubic amplitude in $4D$ with massless particles of arbitrary spin is given by
\begin{align}
\mathcal{A}^{s_{1}, s_{2}, s_{3}}_{h_1, h_2, h_3}= \begin{cases}\langle 1,2\rangle^{h_{3}s_3-h_{1}s_1-h_{2}s_2}\langle 3,1\rangle^{h_{2}s_2-h_{3}s_3-h_{1}s_1}\langle 2,3\rangle^{h_{1}s_1-h_{2}s_2-h_{3}s_3} & \text { when } h_{1}s_1+h_{2}s_2+h_{3}s_3<0 \\ [1,2]^{h_{1}s_1+h_{2}s_2-h_{3}s_3}[3,1]^{h_{3}s_3+h_{1}s_1-h_{2}s_2}[2,3]^{h_{2}s_2+h_{3}s_3-h_{1}s_1} & \text { when } h_{1}s_1+h_{2}s_2+h_{3}s_3>0\end{cases}\label{nonpamp}
\end{align}
The parity even and odd parts are given by the same expression in spinor helicity variables. Interestingly, it can be shown that the epsilon transformation in section \ref{eptr}
keeps the form of amplitude in spinor helicity variables same upto some imaginary number of $i$. In \eqref{nonpamp} there are a total eight independent helicity components $---,--+,-+-,+--$ and their conjugate $+++,++-,+-+,-++$. For each helicity component together with its conjugate, one can construct one parity even and one parity odd amplitude. This implies one can define total four parity even and four parity odd amplitudes in spinor helicity variables. Let us consider a few examples.

\subsection*{spin$_0$-spin$_0$-spin$_s$}
The amplitude for this case is given by
\begin{align}
\mathcal{A}^{00s}_{+} &= \left(g_{ e}-i g_{ o}\right) [1,2]^s[2,3]^{-s}[3,1]^{-s} \quad \mathcal{A}^{00s}_{-} = \left(g_{ e}+i g_{ o}\right)\langle 1,2\rangle^s \langle 2,3\rangle^{-s}\langle 3,1\rangle^{-s} \label{00samp}
\end{align}
which has one parity odd and one parity even amplitude.

\subsection*{graviton-graviton-graviton}
Three graviton amplitude in spinor helicity variables are given by
\begin{align}
\mathcal{A}^{222}_{+++} &= \left(g_{nm, e}-i g_{nm, o}\right) [1,2]^2[2,3]^2[3,1]^2 \quad \mathcal{A}^{222}_{---} = \left(g_{nm, e}+i g_{nm, o}\right)\langle 1,2\rangle^2 \langle 2,3\rangle^2\langle 3,1\rangle^2 \notag\\
\mathcal{A}^{222}_{-++} &= \left(g_{m, e}-i g^{o}_{m}\right)\frac{[2,3]^6}{[1,2]^2[3,1]^2} \quad ~~~~~~~\mathcal{A}^{222}_{+--} = \left(g_{m, e}+i g_{m, o}\right) \frac{\langle 2,3\rangle^6}{\langle 1,2\rangle^2\langle 3,1\rangle^2} \label{gggamp}
\end{align}
where superscript $e,o$ stands for even or odd and subscript $m,nm$ stands for minimal and non-minimal coupling respectively.
Other spinor helicity components can be obtained by permutation.
\subsection*{photon-photon-graviton}
Two photon one graviton amplitude is given by
\begin{align}
&\mathcal{A}^{112}_{+++} =\left(g_{nm, e}-i g_{nm, o}\right)[2,3]^2[3,1]^2 \quad \mathcal{A}^{112}_{---} =\left(g_{nm, e}+i g_{nm, o}\right)\langle 2,3\rangle^2\langle 3,1\rangle^2 \notag\\
&\mathcal{A}^{112}_{-++} =\left(g_{m, e}-i g_{m, o}\right)\frac{[2,3]^4}{[1,2]^2} \quad ~~~~~~~\mathcal{A}^{112}_{+--} =\left(g_{m, e}+i g_{m, o}\right)\frac{\langle 2,3\rangle^4}{\langle 1,2\rangle^2} \nonumber\\
&[\mathcal{A}_{m}]^{112}_{+-+} = \left(g_{m, e}-i g_{m, o}\right)\frac{[3,1]^4}{[1,2]^{2}} \quad [\mathcal{A}_{m}]^{112}_{-+-} = \left(g_{m, e}+i g_{m, o}\right) \frac{\langle 3,1\rangle^{4}}{\langle 1,2\rangle^{2}}
\label{ppgamp}
\end{align}
where the last two lines have identical couplings because of $1 \leftrightarrow 2$ exchange symmetry.
One can also naively add net helicity amplitudes \footnote{
\begin{equation}
\mathcal{A}^{112}_{++-} =\left(g'_{m, e}-i g'_{m, o}\right)\frac{[2,3]^4}{[1,2]^2[3,1]^2} \quad \mathcal{A}^{112}_{--+} = \left(g'_{m, e}+i g'_{m, o}\right)\frac{\langle 2,3\rangle^4}{\langle 1,2\rangle^2\langle 3,1\rangle^2}
\end{equation} }$\mathcal{A}^{112}_{++-}, \mathcal{A}^{112}_{--+}$ but as is argued in Appendix \ref{zeroamp}, they do not lead to any consistent amplitude and hence can be ignored.

\subsection{Cubic vertices in spinor-helicity variables}
From (\cite{Conde:2016izb}), the minimal-coupling vertices with $s_1 \leq s_2 \leq s_3 $ in $4D$ is given by
\begin{align}
&\mathcal{M}^{s_1, s_2, s_3}_{m} = g_{m, e} G^{s_1} Y^{s_2-s_1}_2Y^{s_3-s_1}_3+ g_{nm, o} V_1 G^{s_1} Y^{s_2-s_1-1}_2Y^{s_3-s_1-1}_3 \\
&\mathcal{M}^{s_1, s_2, s_3}_{nm} = g_{nm, e} Y^{s_1}_1 Y^{s_2}_2 Y^{s_3}_3+ g_{nm, o} V_1 Y^{s_1}_1 Y^{s_2-1}_2 Y^{s_3-1}_3
\label{cubvms}
\end{align}
which in the spinor-helicity variables give \footnote{In spinor helicity variables, parity even and parity odd amplitudes are identical up overall factor.}
\small
\begin{align}
&[\mathcal{M}_{nm}]^{s_1, s_2, s_3}_{+++} = g_{A, nm} [1,2]^{s_{1}+s_{2}-s_{3}}[2,3]^{s_{2}+s_{3}-s_{1}}[3,1]^{s_{1}+s_{3}-s_{2}} \nonumber\\ &[\mathcal{M}_{nm}]^{s_1, s_2, s_3}_{---} = g_{H, nm} \langle 1,2\rangle^{s_{1}+s_{2}-s_{3}}\langle 2,3\rangle^{s_{2}+s_{3}-s_{1}}\langle 3,1\rangle^{s_{1}+s_{3}-s_{2}} \notag \\
&[\mathcal{M}_{m}]^{s_1, s_2, s_3}_{-++} = g_{A, m} \frac{[2,3]^{s_{1}+s_{2}+s_{3}}}{[1,2]^{s_{1}+s_{3}-s_{2}}[3,1]^{s_{1}+s_{2}-s_{3}}} \quad [\mathcal{M}_{m}]^{s_1, s_2, s_3}_{+--} = g_{H, m} \frac{\langle 2,3\rangle^{s_{1}+s_{2}+s_{3}}}{\langle 1,2\rangle^{s_{1}+s_{3}-s_{2}}\langle 3,1\rangle^{s_{1}+s_{2}-s_{3}}} \notag \\
&[\mathcal{M}_{m}]^{s_1, s_2, s_3}_{+-+} = g_{A, m} f_{2} \frac{[3,1]^{s_{1}+s_{2}+s_{3}}}{[1,2]^{s_{2}+s_{3}-s_{1}}[2,3]^{s_{1}+s_{2}-s_{3}}} \quad [\mathcal{M}_{m}]^{s_1, s_2, s_3}_{-+-} = g_{H, m} f_{2} \frac{\langle 3,1\rangle^{s_{1}+s_{2}+s_{3}}}{\langle 1,2\rangle^{s_{2}+s_{3}-s_{1}}\langle 2,3\rangle^{s_{1}+s_{2}-s_{3}}} \notag\\
&[\mathcal{M}_{m}]^{s_1, s_2, s_3}_{++-} = g_{A, m} f_{3} \frac{[1,2]^{s_{1}+s_{2}+s_{3}}}{[2,3]^{s_{1}+s_{3}-s_{2}}[3,1]^{s_{2}+s_{3}-s_{1}}} \quad [\mathcal{M}_{m}]^{s_1, s_2, s_3}_{--+} =g_{H, m} f_{3} \frac{\langle 1,2\rangle^{s_{1}+s_{2}+s_{3}}}{\langle 2,3\rangle^{s_{1}+s_{3}-s_{2}}\langle 3,1\rangle^{s_{2}+s_{3}-s_{1}}} \label{cubv}
\end{align}
\normalsize
where
\begin{align}\label{vsph}
&f_{2}=\left(\frac{\langle 1,2\rangle[1,2]\langle 2,3\rangle[2,3]}{\langle 3,1\rangle[3,1]}\right)^{s_{2}-s_{1}}, \quad f_{3}=\left(\frac{\langle 3,1\rangle[3,1]\langle 2,3\rangle[2,3]}{\langle 1,2\rangle[1,2]}\right)^{s_{3}-s_{1}}\\
& g_{A,m} = g_{m, e}+ig_{m, o} \quad g_{H,m} = g_{m, e}-i g_{m, o},~~~g_{A,nm} = g_{nm, e}+i g_{nm, o} \quad g_{H,nm} = g_{nm, e}-i g_{nm, o}.
\end{align}
For the case $s_1\ne s_2 \ne s_3,$ $f_2=f_3=0$ and hence only giving a total four amplitudes. For the special case of $s_1 = s_2 < s_3$, expression for $f_3$ differs slightly from the above and is given by
\begin{align}
&f_{3}=\frac{(\langle 3,1\rangle[3,1])^{s_3-s_1+1}(\langle 2,3\rangle[2,3])^{s_3-s_1}}{(\langle 1,2\rangle[1,2])^{s_3-s_1+1}}
\end{align}
Let us note that $f_2, f_3$ are zero due to momentum conservation, since, $p_i.p_j = \langle ij\rangle [ij] = 0$. This implies many of the components of \eqref{cubv} are zero. For special cases such as $s_1=s_2 \neq s_3$ factor $f_2$ drops out whereas for $s_1=s_2=s_3$ factor $f_2,f_3$ dropout. Also for $s_1=s_2=s_3$ we have $g_{m}^o=0$, that is parity odd minimal cubic vertex does not exist. Here we present some examples.
\subsection*{graviton-graviton-graviton}
\begin{align}\label{222cv}
&[\mathcal{M}_{nm}]^{222}_{+++} = \left(g_{nm, e}+i g_{nm, o}\right) [1,2]^2[2,3]^2[3,1]^2 \quad [\mathcal{M}_{nm}]^{222}_{---} = \left(g_{nm, e}-i g_{nm, o}\right) \langle 1,2\rangle^2 \langle 2,3\rangle^2\langle 3,1\rangle^2 \nonumber\\
&[\mathcal{M}_{m}]^{222}_{-++} = g_{m, e} \frac{[2,3]^6}{[1,2]^2[3,1]^2} \quad [\mathcal{M}_{m}]^{222}_{+--} = g_{m, e} \frac{\langle 2,3\rangle^6}{\langle 1,2\rangle^2\langle 3,1\rangle^2}
\end{align}
and permutations.
We notice that minimal amplitude does not have any parity odd contribution in contrast to \eqref{gggamp}.
\subsection*{photon-photon-graviton}
\begin{align}
&[\mathcal{M}_{nm}]^{112}_{+++} = g_{A, nm} [2,3]^2[1,2]^2 \quad [\mathcal{M}_{nm}]^{112}_{---} = g_{H, nm} \langle 2,3\rangle^2\langle 1,2\rangle^2 \notag \\
&[\mathcal{M}_{m}]^{112}_{-++} = g_{A, m} \frac{[2,3]^{4}}{[1,2]^{2}} \quad ~~~~[\mathcal{M}_{m}]^{112}_{+--} = g_{H, m} \frac{\langle 2,3\rangle^4}{\langle 1,2\rangle^2} \notag \\
&[\mathcal{M}_{m}]^{112}_{+-+} = g_{A, m} \frac{[3,1]^4}{[1,2]^{2}} \quad [\mathcal{M}_{m}]^{112}_{-+-} = g_{H, m} \frac{\langle 3,1\rangle^{4}}{\langle 1,2\rangle^{2}} \label{ppgcubv}
\end{align}
where we have not considered net helicity zero amplitudes\footnote{It is easy to check that they are zero any way for this case.}.
Let us note that for this case minimal and non-minimal amplitude contains both parity even and odd parts.

\subsection{Mismatch between spinor helicity amplitude and covariant vertex}
It turns out that (\ref{cubvms}) doesn't reproduce the full amplitude (\ref{nonpamp}).
The mismatch is most clearly understood for the case of unequal spin. Let us consider an example.
\subsubsection{$s_1\neq s_2 \neq s_3$ with $s_3\ge s_3\ge s_1$}
The covariant amplitude in spinor helicity variables is given by \eqref{cubv}
\small
\begin{align}
&[\mathcal{M}_{nm}]^{s_1, s_2, s_3}_{+++} = g_{A, nm} [1,2]^{s_{1}+s_{2}-s_{3}}[2,3]^{s_{2}+s_{3}-s_{1}}[3,1]^{s_{1}+s_{3}-s_{2}} \nonumber\\ &[\mathcal{M}_{nm}]^{s_1, s_2, s_3}_{---} = g_{H, nm} \langle 1,2\rangle^{s_{1}+s_{2}-s_{3}}\langle 2,3\rangle^{s_{2}+s_{3}-s_{1}}\langle 3,1\rangle^{s_{1}+s_{3}-s_{2}} \notag \\
&[\mathcal{M}_{m}]^{s_1, s_2, s_3}_{-++} = g_{A, m} \frac{[2,3]^{s_{1}+s_{2}+s_{3}}}{[1,2]^{s_{1}+s_{3}-s_{2}}[3,1]^{s_{1}+s_{2}-s_{3}}} \quad [\mathcal{M}_{m}]^{s_1, s_2, s_3}_{+--} = g_{H, m} \frac{\langle 2,3\rangle^{s_{1}+s_{2}+s_{3}}}{\langle 1,2\rangle^{s_{1}+s_{3}-s_{2}}\langle 3,1\rangle^{s_{1}+s_{2}-s_{3}}} \notag \\
&[\mathcal{M}_{m}]^{s_1, s_2, s_3}_{+-+} =0 \quad [\mathcal{M}_{m}]^{s_1, s_2, s_3}_{-+-} =0 \notag\\
&[\mathcal{M}_{m}]^{s_1, s_2, s_3}_{++-} = 0\quad [\mathcal{M}_{m}]^{s_1, s_2, s_3}_{--+} =0 \label{cubv1}
\end{align}
\normalsize

Writing explicitly the spinor helicity amplitude we obtain
\small
\begin{align}
&[\mathcal{A}_{nm}]^{s_1, s_2, s_3}_{+++} = g_{A, nm} [1,2]^{s_{1}+s_{2}-s_{3}}[2,3]^{s_{2}+s_{3}-s_{1}}[3,1]^{s_{1}+s_{3}-s_{2}} \nonumber\\ &[\mathcal{A}_{nm}]^{s_1, s_2, s_3}_{---} = g_{H, nm} \langle 1,2\rangle^{s_{1}+s_{2}-s_{3}}\langle 2,3\rangle^{s_{2}+s_{3}-s_{1}}\langle 3,1\rangle^{s_{1}+s_{3}-s_{2}} \notag \\
&[\mathcal{A}_{m}]^{s_1, s_2, s_3}_{-++} = g_{A, m} \frac{[2,3]^{s_{1}+s_{2}+s_{3}}}{[1,2]^{s_{1}+s_{3}-s_{2}}[3,1]^{s_{1}+s_{2}-s_{3}}} \quad [\mathcal{A}_{m}]^{s_1, s_2, s_3}_{+--} = g_{H, m} \frac{\langle 2,3\rangle^{s_{1}+s_{2}+s_{3}}}{\langle 1,2\rangle^{s_{1}+s_{3}-s_{2}}\langle 3,1\rangle^{s_{1}+s_{2}-s_{3}}} \notag \\
&[\mathcal{A}_{m}]^{s_1, s_2, s_3}_{+-+} = g'_{A, m} \frac{[3,1]^{s_{1}+s_{2}+s_{3}}}{[1,2]^{s_{2}+s_{3}-s_{1}}[2,3]^{s_{1}+s_{2}-s_{3}}} \quad [\mathcal{A}_{m}]^{s_1, s_2, s_3}_{-+-} = g'_{H, m} \frac{\langle 3,1\rangle^{s_{1}+s_{2}+s_{3}}}{\langle 1,2\rangle^{s_{2}+s_{3}-s_{1}}\langle 2,3\rangle^{s_{1}+s_{2}-s_{3}}} \notag\\
&[\mathcal{A}_{m}]^{s_1, s_2, s_3}_{++-} = g''_{A, m} \frac{[1,2]^{s_{1}+s_{2}+s_{3}}}{[2,3]^{s_{1}+s_{3}-s_{2}}[3,1]^{s_{2}+s_{3}-s_{1}}} \quad [\mathcal{A}_{m}]^{s_1, s_2, s_3}_{--+} =g''_{H, m} \frac{\langle 1,2\rangle^{s_{1}+s_{2}+s_{3}}}{\langle 2,3\rangle^{s_{1}+s_{3}-s_{2}}\langle 3,1\rangle^{s_{2}+s_{3}-s_{1}}} \label{cubva}
\end{align}
Comparing \eqref{cubv1} with \eqref{cubva}, we see that there is a clear mismatch between the two. We have introduced $g',g''$ to indicate that they are independent structures. One can convert the spinor helicity amplitude to momentum space variables and show that these extra structures do not correspond to any covariant 4d vertex \footnote{See \cite{Krasnov:2021nsq} for recent discussion on this issue. For the case of spin-s-s-2 amplitude, it was shown that $---,--+$ and their conjugates can be covariantized but $-+-$ and its conjugate helicity components can not be covariantized with the help of symmetric tensors. However they can be covariantized with respect to field variables which originate from twisters. Interestingly, the story is the same in AdS as well. We thank E. Skvortsov for emphasizing this to us.}. We will encounter explicit examples when we consider CFT correlators in spinor helicity variables. We shall see that there are CFT correlators which in the flat space limit reproduce the correct spinor helicity amplitude but do not correspond to any covariant vertex.

Let us consider another example.
\subsubsection{photon-photon-$spin_4$}
The spinor helicity amplitude is given by
\begin{align}
&\mathcal{A}^{114}_{---} = \frac{\langle 12\rangle^6\langle 23\rangle^4}{\langle 12\rangle^4} \quad \mathcal{A}^{114}_{+++} = \frac{[12]^6[23]^4}{[12]^4}\\
&\mathcal{A}^{114}_{--+} = \frac{\langle 12\rangle^6}{\langle 31\rangle^4\langle 23\rangle^4}\quad \mathcal{A}^{114}_{++-} = \frac{[12]^6}{[31]^4[23]^4}\\
&\mathcal{A}^{114}_{+--} = \frac{\langle 23\rangle^6\langle 31\rangle^2}{\langle 12\rangle^4}\quad \mathcal{A}^{114}_{-++} = \frac{[23]^6[31]^2}{[12]^4}\label{pp4amp}.
\end{align}
The cubic vertex in spinor helicity variables is given by
\begin{align}
&[\mathcal{M}_{nm}]^{114}_{+++} = [2,3]^2[1,2]^2 \quad [\mathcal{M}_{nm}]^{114}_{---} = \langle 2,3\rangle^2\langle 1,2\rangle^2 \notag \\
&~~~~~~[\mathcal{M}_{m}]^{114}_{-++} = \frac{[2,3]^{6}[3, 1]^2}{[1,2]^{4}} \quad ~~~~~[\mathcal{M}_{m}]^{114}_{+--} = \frac{\langle 2,3\rangle^{6}\langle 3, 1\rangle^2}{\langle 1,2\rangle^{4}} \notag\\
&~~~~~~[\mathcal{M}_{m}]^{114}_{--+} =[\mathcal{M}_{m}]^{114}_{++-} = 0 \label{pp4cubv}
\end{align}
This implies, this cubic amplitude is not consistent with the full non-perturbative amplitude (\ref{pp4amp}), since, we have $\mathcal{M}^{114}_{--+} \neq \mathcal{A}^{114}_{--+}$, $\mathcal{M}^{114}_{++-} \neq \mathcal{A}^{114}_{++-}$.
\normalsize
The mismatch is summarised in the following Table.
\begin{table}[h!]
\begin{tabular}{||c | c | c ||}
\hline
\multicolumn{3}{|c|}{Covariant vertex vs spinor helicity amplitude}\\
\hline\hline
$\mathbf{s_1 \leq s_2 \leq s_3}$ & \begin{tabular}{@{}c@{}}\bf{Covariant}\\\bf{Cubic vertex} \end{tabular} & \begin{tabular}{@{}c@{}}\bf{Spinor Helicity}\\\bf{Amplitude} \end{tabular} \\ [0.5ex]
\hline
$s_1 = s_2 = 0, s_3 = s$ & 1 even & 1 even and 1 odd \\
\hline
$s_1 = 0, s_2 = s_3 = s$ & 1 even, 1 odd & 1 even and 1 odd \\
\hline
$s_1 = 0, s_2 \ne s_3 $ & 1 even, 1 odd & 2 even, 2 odd \\
\hline
$s_1 = s_2 = s_3 $ & 2 even, 1 odd & 2 even, 2 odd \\
\hline
$s_1 = s_2 = \frac{s_3}{2} $ & 2 even, 2 odd & 2 even, 2 odd \\
\hline
$s_1 = s_2 $ & 2 even, 2 odd & 3 even, 3 odd \\
\hline
$s_1 \ne s_2 \ne s_3$ & 2 even, 2 odd & 4 even , 4 odd\\
\hline
\end{tabular}
\caption{Mismatch between spinor helicity amplitude and covariant vertex}
\label{T4a}
\end{table}

\subsection{Resolving the mismatch?}
Table \ref{T4a} summarizes the mismatch between covariant amplitude and spinor helicity amplitude. In this section, we discuss briefly that demanding covariant or local structure for the four-dimensional vertex can not lead to the resolution of this mismatch.
Let us take an example to illustrate this.
\subsubsection{graviton-graviton-graviton}
The covariant vertex in \eqref{222cv} has no parity odd minimal term whereas amplitude in spinor helicity variables \eqref{gggamp} do get contribution from minimal odd amplitude. However this mis-match can be readily resolved if we consider parity odd minimal amplitudes defined in \eqref{epsliontmin}. If we convert this parity odd minimal amplitude in spinor helicity variables we get
\begin{align}\label{222ab}
&[\mathcal{M}']^{222}_{--+} = i\frac{\langle 12\rangle^6}{\langle 23\rangle^2\langle 31\rangle^2}\quad [\mathcal{M}']^{222}_{++-} = -i\frac{[12]^6}{[23]^2[31]^2}\nonumber\\
&[\mathcal{M}']^{222}_{---} = [\mathcal{M}']^{222}_{+++}=0
\end{align}
which are precisely the missing term in (\ref{222cv}) as compared to \eqref{gggamp}.
Let us emphasize again that minimal odd amplitude defined in \eqref{epsliontmin} can not be converted into a covariant 4D amplitude even though it matches with spinor helicity amplitude. One can show that this is a general feature. In other words, it is easy to convert all the extra spinor helicity amplitudes in \eqref{nonpamp} into momentum space variables and show that all these extra spinor helicity amplitudes lead to a vertex that is not covariant.

Interestingly, in AdS space it was shown in light-cone formalism, see \cite{Metsaev:2018xip}, that one can construct vertex equal in number to amplitude in spinor helicity variables in AdS \cite{Nagaraj:2018nxq}. These extra light-cone vertices do not have any covariant analogue. One can take the flat-space limit as well of the AdS amplitude to connect between light-cone vertex and spinor-helicity amplitudes.
This can be schematically represented by
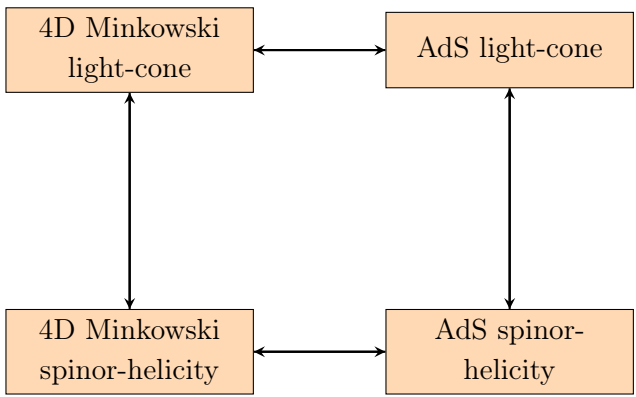
\newline\newline
\begin{tikzpicture}[node distance=2cm]
\node (pro1) [process] {4D Minkowski light-cone};
\node (pro2b) [process, right of=pro1, xshift=3cm] {AdS light-cone};
\node (pro2c) [process, below of=pro1, yshift=-2cm] {4D Minkowski spinor-helicity};
\node (pro2d) [process, below of=pro2b, yshift=-2cm] {AdS spinor-helicity};
\draw [arrow] (pro2b) -- (pro1);
\draw [arrow] (pro1) -- (pro2b);
\draw [arrow] (pro2d) -- (pro2c);
\draw [arrow] (pro2c) -- (pro2d);
\draw [arrow] (pro1) -- (pro2c);
\draw [arrow] (pro2c) -- (pro1);
\draw [arrow] (pro2b) -- (pro2d);
\draw [arrow] (pro2d) -- (pro2b);
\end{tikzpicture}
\newline\newline

We'll see below that we can construct CFT correlator that reproduces correct amplitude in spinor helicity variables but does not correspond to any covariant vertex in the flat space limit.

\section{Flat-space limit of CFT correlators and mismatch with flat space covariant vertex and spinor helicity amplitude}\label{fltsplmt}
This section aims to discuss the general map of the CFT correlation function to the amplitude that we discussed in the previous sections. As we shall demonstrate, in the flat space limit, the CFT correlator obtained in \cite{Bzowski:2013sza,Bzowski:2017poo,Jain:2021gwa,Jain:2021vrv} in general gives rise to flat space covariant amplitude discussed in section \ref{rvw1} but not the full spinor helicity amplitude in \eqref{nonpamp}. In the process, we shall see that there is no analogue of the parity odd minimal amplitude in the case of the CFT correlator.

For a general CFT correlator $\langle J_{s_1}(k_1)J_{s_2}(k_2)J_{s_3}(k_3)\rangle$ we have over all delta function $\delta^3({\vec {k}}_1+{\vec {k}}_2+{\vec {k}}_3)$ where ${\vec {k}}$ is three dimensional momentum vector. However four dimensional amplitude has four dimensional delta function. This missing delta function can be recovered if we set zeroth component of the 4D momentum vector to be $k^0=|{\vec {k}}|$ and work in the limit $E=|{\vec {k_1}}|+|{\vec {k_2}}|+|{\vec {k_3}}|\rightarrow 0.$ We call $E$ to be total energy and it is clear that setting $E\rightarrow 0$ provides us with the extra delta function \footnote{There are other ways to get the flat space amplitude starting from CFT correaltion function. For example, see \cite{Li:2021snj} for a recent discussion on the flat-space limit of the CFT correlator, where one of the scaling dimensions of the CFT correlator was taken to be large.}. For simplicity of notation we simply denote $ |{\vec {k_i}}|$ simply by $k_i.$

\subsection{Flat space limit of CFT correlator of exactly conserved currents: No parity odd minimal analogue}
CFT correlator of conserved currents can be split up into homogeneous and non-homogeneous pieces as follows \cite{Jain:2021vrv}
\begin{equation}
\langle J_{s_1} J_{s_2} J_{s_3} \rangle = \langle J_{s_1} J_{s_2} J_{s_3} \rangle_{h}+ \langle J_{s_1} J_{s_2} J_{s_3} \rangle_{nh}.
\end{equation}
These satisfy following conformal ward identity
\begin{align}\label{cwt1}
&\widetilde{K}^{\kappa}\left\langle\frac{J_{s_{1}}}{k_{1}^{s_{1}-1}} \frac{J_{s_{2}}}{k_{2}^{s_{2}-1}} \frac{J_{s_{3}}}{k_{3}^{s_{3}-1}}\right\rangle_{nh}=\text { Ward identity}\nonumber\\
&\widetilde{K}^{\kappa}\left\langle\frac{J_{s_{1}}}{k_{1}^{s_{1}-1}} \frac{J_{s_{2}}}{k_{2}^{s_{2}-1}} \frac{J_{s_{3}}}{k_{3}^{s_{3}-1}}\right\rangle_{h}=0
\end{align}
where $\widetilde{K}^{\kappa}$ is a special conformal generator. It was further shown in \cite{Gandhi:2021gwn,Jain:2021gwa} that homogeneous and non-homogeneous pieces can further be identified as
\begin{align}
\langle J_{s_1} J_{s_2} J_{s_3} \rangle_{h} =\frac{ \langle J_{s_1} J_{s_2} J_{s_3} \rangle_{FB}- \langle J_{s_1} J_{s_2} J_{s_3} \rangle_{FF}}{2}\nonumber\\
\langle J_{s_1} J_{s_2} J_{s_3} \rangle_{nh} =\frac{ \langle J_{s_1} J_{s_2} J_{s_3} \rangle_{FB}+ \langle J_{s_1} J_{s_2} J_{s_3} \rangle_{FF}}{2}
\end{align}
where FB and FF stand for free bosonic and free fermionic theory respectively. It was shown in \cite{Giombi:2011rz} that for exactly conserved currents satisfying triangle inequality $s_i\le s_j+s_k$, there is also a parity odd contribution. It was shown in \cite{Jain:2021vrv} that parity odd contribution is always homogeneous and further it was shown in \cite{Jain:2021gwa} that the parity odd term can be calculated using free bosonic and free fermion answer using some epsilon transformation as follows
\begin{align}\label{s1s2s3g}
\langle J_{s_1} J_{s_2}J_{s_3}\rangle_{h,o} &=[\mathcal{O}_\epsilon]\frac{\langle J_{s_1} J_{s_2}J_{s_3}\rangle_{FB} -\langle J_{s_1} J_{s_2}J_{s_3}\rangle_{FF} }{2}.
\end{align}
In the last line of \eqref{s1s2s3g} we have used the fact that epsilon transform of parity-even H term produces parity-odd H term, see \cite{Jain:2021gwa} for more details.
In the flat-space limit one obtains the following map
\begin{align}\label{mapamp1}
\langle J_{s_1} J_{s_2}J_{s_3}\rangle_{nh,e}&\rightarrow M^{s_1 s_2 s_3}_{m,e}\nonumber\\
\langle J_{s_1} J_{s_2}J_{s_3}\rangle_{h,e}&\rightarrow M^{s_1 s_2 s_3}_{nm,e}\nonumber\\
\langle J_{s_1} J_{s_2}J_{s_3}\rangle_{h,o}&\rightarrow M^{s_1 s_2 s_3}_{nm,o}
\end{align}
This implies that parity-odd minimal amplitude in section \ref{rvw1} is not reproduced by CFT correlators of conserved currents in the flat space limit.
When the spin violates the triangle inequality $s_i > s_j+s_k,$ as was shown in \cite{Jain:2021whr} the only contribution to correlation function comes from non-homogeneous pieces. Furthermore, for exactly conserved currents there is no parity-odd contribution for this configuration. One obtains the following map in this case
\begin{align}\label{s1s2s3gout}
\langle J_{s_1} J_{s_2}J_{s_3}\rangle_{nh, e, 1} &=\frac{\langle J_{s_1} J_{s_2}J_{s_3}\rangle_{FB} +\langle J_{s_1} J_{s_2}J_{s_3}\rangle_{FF} }{2}\rightarrow M^{s_1 s_2 s_3}_{m,e} \nonumber\\
\langle J_{s_1} J_{s_2}J_{s_3}\rangle_{nh, e, 2} &=\frac{\langle J_{s_1} J_{s_2}J_{s_3}\rangle_{FB} -\langle J_{s_1} J_{s_2}J_{s_3}\rangle_{FF} }{2}\rightarrow M^{s_1 s_2 s_3}_{nm,e}
\end{align}
It is very clear that, there is a mismatch with CFT correlator with flat space covariant vertex as well as with flat space spinor helicity amplitude. We summarise the mismatch in the following table.
\newpage
\begin{table}[h!]
\begin{tabular}{||c | c | c | c ||}
\hline
\multicolumn{4}{|c|}{Mismatch between amplitude and CFT correlator}\\
\hline\hline
$\mathbf{s_1 \leq s_2 \leq s_3}$ & \begin{tabular}{@{}c@{}}\bf{Covariant}\\\bf{Cubic vertex} \end{tabular} & \begin{tabular}{@{}c@{}}\bf{Spinor Helicity}\\\bf{Amplitude} \end{tabular} & \begin{tabular}{@{}c@{}}\bf{CFT Correlator}\\\bf{of conserved currents} \end{tabular}\\ [0.5ex]
\hline
$s_1 = s_2 = 0, s_3 = s$ & 1 even & 1 even and 1 odd & 1 even \\
\hline
$s_1 = 0, s_2 = s_3 = s$ & 1 even, 1 odd & 2 even, 2 odd & 1 even, 1 odd \\
\hline
$s_1 = 0, s_2 \ne s_3 $ & 1 even, 1 odd & 2 even, 2 odd & 1 even, 1 odd \\
\hline
$s_1 \leq s_2 \leq s_3$ with $s_3\le s_1+s_2$ & 2 even, 2 odd & 4 even , 4 odd & 2 even, 1 odd\\
\hline
$s_1 \leq s_2 \leq s_3$ with $s_ 3> s_1+s_2$ & 2 even, 2 odd & 4 even , 4 odd & 2 even \\
\hline
\end{tabular}
\caption{Mismatch between spinor helicity amplitude and covariant vertex and CFT correlator.}
\label{msmtchcftamp}
\end{table}
The parity-odd part for $s_3> s_1+s_2$ is generated when we consider the correlation function of slightly-broken higher spin currents.
\subsection*{CFT correlator for weakly broken HS current and flat space limit}
When we consider $s_3 > s_1+s_2,$ with $s_3>s_1\ge s_2,$ the parity even part of CFT correlator and its flat space limit is considered in \eqref{s1s2s3gout}. The parity odd part is non-zero for weakly broken HS current and is given by
\begin{align}\label{goutnh}
\langle J_{s_1} J_{s_2}J_{s_3}\rangle_{nh,o} &=[\mathcal{O}_\epsilon]\frac{\langle J_{s_1} J_{s_2}J_{s_3}\rangle_{FB} -\langle J_{s_1} J_{s_2}J_{s_3}\rangle_{FF} }{2}.
\end{align}
One obtains the following map
\begin{align}\label{oddout}
\langle J_{s_1} J_{s_2}J_{s_3}\rangle_{nh,o}\rightarrow M^{s_1 s_2 s_3}_{nm,o}.
\end{align}
Explicit examples are worked out in the appendix \ref{CFTfta}.
In the following section, we construct a parity-odd CFT correlator which in the flat-space limit produces correct parity-odd minimal amplitude.
Correlation function involving one or more scalars maps trivially to the flat-space limit. For details see Appendix \ref{CFTfta}. Let us note that there is no analogue of parity-odd minimal amplitude in CFT as can be seen in \eqref{mapamp1}, \eqref{oddout} when we consider exactly conserved currents or weakly broken currents. In the next section, we construct such a parity odd CFT correlator which in the flat space limit reproduces missing parity odd covariant amplitude.
\section{CFT Correlator/ covariant vertex correspondence: A new parity-odd CFT correlation function}\label{nptod}
In this section, we construct parity-odd CFT correlation function which in the flat-space limit goes over to parity-odd minimal amplitude listed in section \ref{rvw1}. In the case of flat-space amplitude, we explicitly showed in section \ref{eptr} that the parity-odd minimal amplitude is obtained from parity-even minimal amplitude by doing what is called epsilon transform. Our proposal for the CFT correlation function is exactly the analogue of the amplitude case.

We propose
\begin{align}\label{goutnh1}
\langle J_{s_1} J_{s_2}J_{s_3}\rangle_{nh,o}=[\mathcal{O}_\epsilon] \langle J_{s_1} J_{s_2}J_{s_3}\rangle_{nh,e} &=[\mathcal{O}_\epsilon]\frac{\langle J_{s_1} J_{s_2}J_{s_3}\rangle_{FB} +\langle J_{s_1} J_{s_2}J_{s_3}\rangle_{FF} }{2}
\end{align}
which by construction produces the correct flat-space minimal parity-odd amplitude\footnote{In spinor helicity variables, the parity even non-homogeneous term used in \eqref{goutnh1} and parity odd term defined in the same equation \eqref{goutnh1} are identical upto imaginary factor of $i.$}.
Since this is non-homogeneous, this CFT correlator satisfies Ward-Takahashi (WT) identity and is given by
\begin{align}\label{goutnh2}
WT[\langle J_{s_1} J_{s_2}J_{s_3}\rangle_{nh,o}] &=[\mathcal{O}_\epsilon]\frac{WT_{FB}+WT_{FF}}{2}.
\end{align}
When the spins satisfy triangle inequality $s_i\le s_j+s_k,$ we have $WT_{FB}=WT_{FF}$ \cite{Jain:2021whr} which gives
\begin{align}\label{goutnh3}
WT[\langle J_{s_1} J_{s_2}J_{s_3}\rangle_{nh,o}] &=[\mathcal{O}_\epsilon]WT_{FB}.
\end{align}
For correlators with spin which violates the triangle inequality, we have $WT_{FB}\neq WT_{FF}$ and hence we have to use \eqref{goutnh1}.
The explicit form of WT identity for general spins can be very complicated. Below we work out a few simple examples of this correlator \eqref{goutnh1} and their flat-space limit.
We also discuss the WT identity \eqref{goutnh2}.
\subsection{Example: $\langle JJT\rangle $}
Let us consider the simplest example of two spin-1 and one spin-2 operators.
The parity even part of the CFT correlator is given by \cite{Bzowski:2013sza,Jain:2021vrv}
\begin{align}
&\langle J(z_1, k_1)J(z_2, k_2)T(z_3, k_3)\rangle_{e} =\langle J(z_1, k_1)J(z_2, k_2)T(z_3, k_3)\rangle_{h,e} +\langle J(z_1, k_1)J(z_2, k_2)T(z_3, k_3)\rangle_{nh,e} \nonumber\\ &=\frac{2(3k_3+E)}{E^4}(z_3.k_2)^2(z_2.k_3)(z_1.k_3)+\left[\frac{2k^2_3}{E^3}-c_J\frac{2(k_3+E)}{E^2}\right](z_3.k_2)^2(z_1.z_2)\notag\\&+\left[ \frac{1}{E^3}(-2k^2_3-k^2_2+k^2_1-3k_3k_2+3k_3k_1)-c_J\frac{2(k_3+E)}{E^2}\right](z_3.z_2)(z_3.k_2)(z_1.k_3)\notag\\&+\left[ \frac{1}{E^3}(-2k^2_3-k^2_1+k^2_2-3k_3k_1+3k_3k_2)-c_J\frac{2(k_3+E)}{E^2}\right](z_1.z_3)(z_3.k_2)(z_2.k_1)\notag\\&+\left[\frac{(k_3+E)(k^2_3-(k_2+k_1)^2+4 k_2 k_1)}{2E^2}+c_J (\frac{2k^2_3}{E}-k_2-k_1)\right](z_1.z_3)(z_3.z_2)
\end{align}
where term proportional to $c_J$ is non-homogeneous parity even part and rest is parity even homogeneous contribution. In the flat-space limit we get
\begin{align}
&\lim_{E\to 0}\langle J(z_1, k_1)J(z_2, k_2) T(z_3, k_3)\rangle = \frac{6k_3}{E^4}(z_3.k_2)^2(z_2.k_3)(z_1.k_2)+O(\frac{1}{E^3})\notag\\&+c_J\frac{2k^2_3}{E^3}(z_1.z_2 z_3.k_1+z_2.z_3 z_1.k_2+z_3.z_1 z_2.k_3)(z_3.k_2) +c_J ~O(\frac{1}{E^2})
\end{align}
which perfectly match with parity even minimal and non-minimal vertices
\begin{align}
\mathcal{M}_{nm, e} = (z_3.k_1)^2(z_2.k_3)(z_1.k_2) \quad \mathcal{M}_{m, e} = (z_1.z_2 z_3.k_1+z_2.z_3 z_1.k_2+z_3.z_1 z_2.k_3)(z_3.k_2).
\end{align}
The non-minimal parity odd amplitude is obtained by taking the flat space limit of parity odd homogeneous contribution. The parity odd CFT correlator can be found in \cite{Jain:2021vrv}. It is easy to show that in the flat space limit
\begin{align}
\lim_{E\rightarrow 0}\langle JJT\rangle _{h,o}\rightarrow \mathcal{M}_{nm, o}.
\end{align}
However as it is clear, there is no analogue of the CFT correlator which in the flat space limit reproduces correct flat space minimal parity odd amplitude.

We now show using definition \eqref{goutnh1} we get a CFT correlator which in the flat space reproduces correct parity odd minimal amplitude. As was discussed below \eqref{sss3}, in this case also we need to introduce Chan-Paton factors and anti symmetric with respect to two spin-1 currents to get parity-odd non-homogeneous results.
Using the proposal (\ref{goutnh1}), we see that for $\langle TJJ\rangle_{nh, odd}$, we get
\begin{align}
\langle JJT\rangle_{nh, o} &= ([O_{\epsilon}]_2-[O_{\epsilon}]_1)\langle JJT\rangle_{nh,e} \nonumber\\
&=([O_{\epsilon}]_2-[O_{\epsilon}]_1)\frac{\langle JJT\rangle_{FB}+\langle JJT\rangle_{FF}}{2} \nonumber\\
&= A(k_1, k_2, k_3)z_3.k_1z_3.z_2\epsilon(z_1k_1k_3) + B(k_1, k_2, k_3)z_3.z_2\epsilon(z_1 k_1 z_3)\notag\\&+C(k_1, k_2, k_3)z_3.k_1 z_2.k_3\epsilon(z_1 k_1 z_3)+D(k_1, k_2, k_3)(z_3.k_1)^2\epsilon(z_1k_1z_2)\notag\\&-(2\leftrightarrow 1)\label{TJJnho}
\end{align}
where
\begin{align}
D=C=-A = \frac{E+k_3}{k_2E^2}\quad B = -\frac{1}{2k_2}\left(-k_2-k_1+\frac{2k^2_3}{E}\right).
\end{align}
This new parity odd CFT correlator has a pole only in the total energy $E.$
The ward identity can be obtained using the proposal (\ref{goutnh2})
\begin{align}
\langle JJk_3.T\rangle_{nh, o} &= ([O_{\epsilon}]_2-[O_{\epsilon}]_1)\langle JJk_3.T\rangle_{nh,e} \notag\\&= -z_1.z_3\epsilon(z_2k_2k_3)+\frac{k_1}{k_2}[(z_1.k_2)\epsilon(z_2k_2z_3)-(z_3.k_2)\epsilon(z_2k_2z_1)]-z_3.k_1\epsilon(z_2k_2z_1)\notag\\&-(2\leftrightarrow 1)\label{TJJnhwi}
\end{align}
One can confirm that (\ref{TJJnho}) and (\ref{TJJnhwi}) are consistent with each other by going to Spinor-Helicity variables and checking conformal ward identity.

In the flat-space limit we obtain
\begin{align}
\lim_{E\rightarrow 0} \langle JJT\rangle_{nh, o} \sim \frac{1}{E^{2}} \left(-\left(\epsilon\left(z_{2} z_{3} k_{2}\right) k_{3}-\epsilon\left(z_{2} z_{3} k_{3}\right) k_{2}\right)\left(z_{3} \cdot z_{1}\right)+\epsilon\left(z_{1} z_{2} z_{3}\right) k_{2}\left(z_{3} \cdot k_{1}\right)\right)+{\mathcal{O}}(\frac{1}{E})
\end{align}
which matches with minimal odd amplitude in \eqref{JJT3D} and converting it in 4D notation we obtain amplitude given in \eqref{evamp1}.



Even though the correlator that has been constructed has a nice behaviour that it has only total energy singularity, it also has some other unusual properties which become clear in position space.
\subsubsection{In position space}
In position space, the epsilon transform is given by
\begin{align}
[O_{\epsilon}]_{I}: \langle O_1(x_1)\cdots O^{\mu_1\cdots\mu_i\cdots\mu_I}(x_I)\cdots O_n(x_n)\rangle \to \epsilon^{\mu_i}_{~~\sigma\alpha}\int\frac{d^3y_I}{|x_I-y_I|^2}\partial^{\sigma}_{x_I}\langle O_1(x_1)\cdots O^{\mu_1\cdots\alpha\cdots\mu_I}(y_I)\cdots O(x_n)\rangle\label{epps}
\end{align}
From the above, it is clear that the transformation is not so straightforward in position space, unlike the momentum space.

Let us look into the simplest example $\langle TJJ\rangle.$ To start with, let us consider the epsilon transform of ward identity. The ward identity for $\langle TJJ\rangle$ in position space is given by
\begin{align}\label{wtjjt1}
\partial_3^{\mu}\langle J_{\rho}(x_1)J_{\sigma}(x_2)T_{\mu\nu}(x_3)\rangle &= \partial_{3\nu}\delta^{(3)}(x_3-x_2)\langle J_{\sigma}(x_3)J_{\rho}(x_1)\rangle-\partial_{3\mu}\delta^{(3)}(x_3-x_2)\delta_{\nu\sigma}\langle J_{\mu}(x_3)J_{\rho}(x_1)\rangle\notag\\&+\partial_{\nu}\delta^{(3)}(x_1-x_3)\langle J_{\sigma}(x_2) J_{\rho}(x_3)\rangle-\partial_{\mu}\delta^{(3)}(x_1-x_3)\delta_{\nu\rho}\langle J_{\sigma}(x_2)J_{\mu}(x_1)\rangle
\end{align}
Using (\ref{epps}) in the above equation, we get
\begin{align}\label{anhwt1}
&\partial_3^{\mu}\langle J_{\rho}(x_1)J_{\sigma}(x_2)T_{\mu\nu}(x_3)\rangle_{nh, o}
= ([O_{\epsilon}]_2-[O_{\epsilon}]_1)\partial_3^{\mu}\langle J_{\rho}(x_1)J_{\sigma}(x_2)T_{\mu\nu}(x_3)\rangle \notag\\&= \epsilon_{\sigma\alpha\tau}\partial_{2\nu}\partial_2^{\alpha}\frac{1}{|x_2-x_3|^2}\langle J^{\tau}(x_3)J^{\rho}(x_1)\rangle-\epsilon_{\sigma\alpha\nu}\partial_{2\mu}\partial^{\alpha}_2\frac{1}{|x_2-x_3|^2}\langle J^{\mu}(x_3)J^{\rho}(x_1)\rangle\notag\\&+\partial_{3\nu}\delta^{(3)}(x_1-x_3)\langle J^{\sigma}(x_2)J^{\rho}(x_3)\rangle_{o}-\partial_{1\mu}\delta^{(3)}(x_1-x_3)\delta_{\nu\rho}\langle J^{\sigma}(x_2)J^{\mu}(x_3) \rangle_{o}\notag\\&-[(2, \sigma)\leftrightarrow (1, \rho)]
\end{align}
In the first line, we made use of the properties of the delta function and we also
have made use of
\begin{align}
\left\langle J_{\mu}(x) J_{\nu}(y)\right\rangle_{\text {o}}=\epsilon_{\mu \sigma \alpha} \int \frac{d^{3} x_{1}}{\left|x-x_{1}\right|^{2}} \partial_{x_{1}}^{\sigma}\left\langle J^{\alpha}\left(x_{1}\right) J_{\nu}(y)\right\rangle_{\text {e}}
\end{align}
in the second line. The RHS of (\ref{anhwt1}) gives
\begin{align}\label{nhwt10}
&\partial_3^{\mu}\langle J_{\rho}(x_1)J_{\sigma}(x_2)T_{\mu\nu}(x_3)\rangle_{nh, o}
\notag\\&= \epsilon_{\sigma\alpha\tau}\left(-\frac{2\delta^{\alpha}_{\nu}}{x^4_{32}}+\frac{8x^{\alpha}_{32}x_{\nu 32}}{x^6_{32}}\right)\left(\frac{\delta^{\tau}_{\rho}}{x^4_{13}}-\frac{2x^{\tau}_{13}x_{13\rho}}{x^6_{13}}\right)-\epsilon_{\sigma\alpha\nu}\left(-\frac{2\delta^{\alpha}_{\mu}}{x^4_{32}}+\frac{8x^{\alpha}_{32}x_{\mu 32}}{x^6_{32}}\right)\left(\frac{\delta^{\mu}_{\rho}}{x^4_{13}}-\frac{2x^{\mu}_{13}x_{13\rho}}{x^6_{13}}\right)\notag\\&+\partial_{1\nu}\delta^{(3)}(x_1-x_3)\epsilon_{\sigma\rho\tau}\partial^{\tau}_2\delta^{(3)}(x_2-x_3)-\partial_{1\mu}\delta^{(3)}(x_1-x_3)\delta_{\nu\rho}\epsilon_{\sigma\mu\tau}\partial^{\tau}_2\delta^{(3)}(x_2-x_3)\notag\\&-[(2, \sigma)\leftrightarrow (1, \rho)]\nonumber\\
&\equiv \epsilon_{\sigma\alpha\tau}\left(-\frac{2\delta^{\alpha}_{\nu}}{x^4_{32}}+\frac{8x^{\alpha}_{32}x_{\nu 32}}{x^6_{32}}\right)\left(\frac{\delta^{\tau}_{\rho}}{x^4_{13}}-\frac{2x^{\tau}_{13}x_{13\rho}}{x^6_{13}}\right)-\epsilon_{\sigma\alpha\nu}\left(-\frac{2\delta^{\alpha}_{\mu}}{x^4_{32}}+\frac{8x^{\alpha}_{32}x_{\mu 32}}{x^6_{32}}\right)\left(\frac{\delta^{\mu}_{\rho}}{x^4_{13}}-\frac{2x^{\mu}_{13}x_{13\rho}}{x^6_{13}}\right)\notag\\
&-[(2, \sigma)\leftrightarrow (1, \rho)]
\end{align}
where in the last line we have removed contact terms. These contact terms will give rise to a contact term in correlation function, see \cite{Jain:2021vrv} for similar discussion.
However, it is important to note that, unlike in \eqref{wtjjt1}, no delta function appears in \eqref{nhwt10}. This implies that spin-2 current is not conserved even away from coincident points. This implies we can't identify this spin-2 current as a stress tensor. This is as expected as for exactly conserved current we can't have more than three structures.
\subsection*{WT identity for $\langle J T J_3\rangle$}
The fact that Ward-Takahashi identity is non zero even away from contact points, is a universal fact for non-homogeneous parity-odd terms, which can be checked easily. Consider now the ward identity for $\langle J_1 J_2 J_3\rangle$
\begin{align}
&\partial^{\gamma}_3 \langle J_{1\mu}(x_1) T_{\nu\rho}(x_2) J_{3\alpha\beta\gamma}(x_3)\rangle \sim \partial_{3\mu}\delta^{(3)}(x_3-x_1)\langle T_{\alpha\beta}(x_3)T_{\nu\rho}(x_2) \rangle+\partial_{3(\alpha}\delta^{(3)}(x_3-x_1)\langle T_{\beta)\mu}(x_3)T_{\nu\rho}(x_2) \rangle\notag\\&+(3\partial_{3\alpha}\partial_{3\beta}-\delta_{\alpha\beta}\Box_3)\partial_{3(\nu}\delta^{(3)}(x_3-x_2)\langle J_{\rho)}(x_3) J_{1\mu}(x_1) \rangle+(3\partial_{3\nu}\partial_{3\rho}-\delta_{\nu\rho}\Box_3)\partial_{3(\alpha}\delta^{(3)}(x_3-x_2)\langle J_{\beta)}(x_3) J_{1\mu}(x_1) \rangle
\end{align}
After an epsilon transform we get
\begin{align}\label{123wt}
&[O_{\epsilon}]_1\partial^{\gamma}_3 \langle J_{1\mu}(x_1) T_{\nu\rho}(x_2) J_{3\alpha\beta\gamma}(x_3)\rangle \sim \partial_{1\mu}\frac{1}{|x_{13}|^2}\langle T_{\alpha\beta}(x_3)T_{\nu\rho}(x_2) \rangle+\partial_{1(\alpha}\frac{1}{|x_{13}|^2}\langle T_{\beta)\mu}(x_3)T_{\nu\rho}(x_2) \rangle\notag\\&+(3\partial_{3\alpha}\partial_{3\beta}-\delta_{\alpha\beta}\Box_3)\partial_{3(\nu}\delta^{(3)}(x_3-x_2)\langle J_{\rho)}(x_3) J_{1\mu}(x_1) \rangle_{o}+(3\partial_{3\nu}\partial_{3\rho}-\delta_{\nu\rho}\Box_3)\partial_{3(\alpha}\delta^{(3)}(x_3-x_2)\langle J_{\beta)}(x_3) J_{1\mu}(x_1) \rangle_{o}
\end{align}
where
\begin{align}
\langle J_{\alpha)}(x_1) J_{1\beta}(x_2) \rangle_{o} = \epsilon_{\alpha\beta\lambda}\partial^{\lambda}_{1}\delta^{(3)}(x_1-x_2)
\end{align}
Just like $\langle JJT\rangle$, we see that an epsilon transform at $x_1$ gives rise to contact terms in the second line of \eqref{123wt} and is therefore dropped. We also get terms that do not vanish at non-coincident points. Hence, we see that the epsilon transform of conserved current correlations gives rise to correlations that are not conserved.
Let us consider another example, all equal spin $\langle TTT\rangle$.
\subsection{$\langle TTT\rangle$}
Consider now the $\langle TTT\rangle_{e}$ correlator in momentum space
\begin{align}
&\langle T(z_1, k_1)T(z_2, k_2)T(z_3, k_3)\rangle_{e} =c_1 \frac{k_1k_2k_3}{E^6}[2 z_1.k_2 z_2.k_3 z_3.k_1 + E (k_3 z_1.z_2 z_3.k_1 + k_1 z_2.z_3 z_1.k_2 + k_2 z_3.z_1 z_2.k_3)] \notag\\&+ c_T\left(\frac{k_1k_2k_3}{E^2}+\frac{k_1 k_2 +k_2 k_3+ k_3 k_1}{E}-E\right) (z_1.z_2 z_3.k_1 + z_2.z_3 z_1.k_2 + z_3.z_1 z_2.k_3)^2
\end{align}
which in the flat-space limit gives
\begin{align}
&\lim_{E\to 0}\langle T(z_1, k_1)T(z_2, k_2)T(z_3, k_3)\rangle_{e} = c_1\frac{k_1k_2k_3}{E^6}[2 z_1.k_2 z_2.k_3 z_3.k_1 ] +O(\frac{1}{E^5})\notag\\&+ c_T\frac{k_1k_2k_3}{E^2} (z_1.z_2 z_3.k_1 + z_2.z_3 z_1.k_2 + z_3.z_1 z_2.k_3)^2 + c_T O(\frac{1}{E})
\end{align}
which perfectly matches the minimal and non-minimal parity even vertices in (\ref{222cv1}). For parity-odd case \footnote{Following \cite{Jain:2021vrv}, one can write parity-odd non-homogeneous piece as follows
\begin{align}\label{contnhod}
&\langle TTT\rangle_{nh, o} = \frac{1}{24}[\epsilon(z_1z_2k_1)(z_1.z_2)(z_3.k_1)^2-\epsilon(z_1z_2k_2)(z_1.z_3)(z_2.z_3)]\notag\\&+\frac{1}{12}[(z_1.z_3)(z_2.z_3)\epsilon(z_1z_2k_1)(k^2_1+\frac{7}{4}k^2_2+\frac{7}{4}k^2_3)-(z_1.z_2)(z_3.k_1)^2\epsilon(z_1z_2k_2)(k^2_2+\frac{7}{4}k^2_1+\frac{7}{4}k^2_3)]+\text{cyclic terms}
\end{align}
However, one can see that $\langle TTT\rangle_{nh, o}$ is just a contact term. This does not correspond to any cubic vertex.}, we have
\begin{align}
& \langle T(z_1, k_1)T(z_2, k_2)T(z_3, k_3)\rangle_{h, o} \notag\\&= \left(k_{1} k_{2} k_{3}\right) \frac{1}{E^{3}}\left[\left\{\left(\vec{k}_{1} \cdot \vec{z}_{3}\right)\left(\epsilon^{k_{3} z_{1} z_{2}} k_{1}-\epsilon^{k_{1} z_{1} z_{2}} k_{3}\right)+\left(\vec{k}_{3} \cdot \vec{z}_{2}\right)\left(\epsilon^{k_{1} z_{1} z_{3}} k_{2}-\epsilon^{k_{2} z_{1} z_{3}} k_{1}\right)\right.\right.\notag \\
&\left.\left.\quad-\left(\vec{z}_{2} \cdot \vec{z}_{3}\right) \epsilon^{k_{1} k_{2} z_{1}} E+\frac{k_{1}}{2} \epsilon^{z_{1} z_{2} z_{3}} E\left(E-2 k_{1}\right)\right\}+\text { cyclic perm }\right] \notag\\
&\times\left[\frac{1}{E^{3}}\left\{2\left(\vec{z}_{1} \cdot \vec{k}_{2}\right)\left(\vec{z}_{2} \cdot \vec{k}_{3}\right)\left(\vec{z}_{3} \cdot \vec{k}_{1}\right)+E\left\{k_{3}\left(\vec{z}_{1} \cdot \vec{z}_{2}\right)\left(\vec{z}_{3} \cdot \vec{k}_{1}\right)+\text { cyclic }\right\}\right\}\right]
\end{align}
In the flat-space limit, which becomes
\begin{align}
\lim_{E\to 0}\langle TTT\rangle_{h, o} &= \frac{k_1k_2k_3}{E^6}\left[\left(\vec{k}_{1} \cdot \vec{z}_{3}\right)\left(\epsilon^{k_{3} z_{1} z_{2}} k_{1}-\epsilon^{k_{1} z_{1} z_{2}} k_{3}\right)+\text { cyclic perm }\right]\left(\vec{z}_{1} \cdot \vec{k}_{2}\right)\left(\vec{z}_{2} \cdot \vec{k}_{3}\right)\left(\vec{z}_{3} \cdot \vec{k}_{1}\right)
\end{align}
which is precisely the non-minimal parity-odd cubic vertex mentioned in (\ref{222cv1}).
Since the three-point function of conserved currents at maximum can only have three structures, we see that just like cubic vertex we do not have any analogue of parity odd-minimal amplitude at the level CFT correlation function.

However, let us define another parity-odd structure namely
\begin{align}
\langle TTT\rangle'_{o} &= ([O_\epsilon]_1+[O_\epsilon]_2+[O_\epsilon]_3)\langle TTT\rangle_{nh, e} \notag\\&= \frac{E^3-E \left( k_1 k_2 + k_2 k_3 + k_3 k_1\right)-k_1 k_2 k_3}{E^2 k_1 k_2 k_3}(z_1.k_2 z_2.z_3 +z_1.z_2 z_3.k_1+z_2.k_3 z_3.z_1)\notag\\&[k_2 k_3 z_2.z_3 \epsilon(z_1k_1k_2) + \text{cyclic terms}- k_1 k_2 k_3 E \epsilon(z_1z_2z_3)]\label{anoddttt}
\end{align}
which in the flat-space limit gives
\begin{align}
\lim_{E\to 0}\langle TTT\rangle'_{o} \sim -\frac{c_{123}}{E^2}[V']^{222}+{\mathcal{O}}(\frac{1}{E})
\end{align}
which is precisely what we computed in (\ref{epsliontmin}). As mentioned before, this flat-space limit cannot be recast as a $4D$ flat-space amplitude. It is easy to show that, spin two current that appears in \eqref{anoddttt} is not conserved. To show this we work in position space.

\subsubsection{ Epsilon transform WT identity of $\langle TTT\rangle$ in position space}
Consider the $\langle TTT\rangle$ ward identity in position space
\begin{align}
\partial^{\mu}\langle T_{\mu\nu}(x)T_{\sigma\rho}(y) T_{\alpha\beta}\rangle &= \partial_{\nu}\delta^{(3)}(x-y)\langle T_{\sigma\rho}(x)T_{\alpha\beta}(z)\rangle + \left\{\partial_{\sigma}(\delta^{(3)}(x-y)\langle T_{\rho\nu}(x)T_{\alpha\beta}(z)\rangle)+\sigma\leftrightarrow \rho\right\}\notag\\
&+\partial_{\nu}\delta^{(3)}(x-z)\langle T_{\sigma\rho}(y)T_{\alpha\beta}(x)\rangle+\left\{\partial_{\alpha}(\delta^{(3)}(x-z)\langle T_{\beta\nu}(x)T_{\sigma\rho}(y)\rangle)+\alpha\leftrightarrow\beta\right\}
\end{align}
where we now perform an epsilon transform and just like for the case of $\langle JJT\rangle$ we find that
\begin{align}
&[O_{\epsilon}]_y\partial^{\mu}\langle T_{\mu\nu}(x)T_{\sigma\rho}(y) T_{\alpha\beta}\rangle \notag\\&\sim \epsilon_{\sigma\eta\zeta}\partial^{\zeta}\partial_{\nu}\frac{1}{|x-y|^2}\langle T_{\eta\rho}(x)T_{\alpha\beta}(z)\rangle + \partial_{\rho}(\epsilon_{\sigma\eta\zeta}\partial^{\zeta}\frac{1}{|x-y|^2}\langle T_{\eta\nu}(x)T_{\alpha\beta}(z)\rangle)\notag\\
&+\epsilon_{\sigma\eta\zeta}\partial^{\zeta}\frac{1}{|x-y|^2}\partial_{\eta}\langle T_{\rho\nu}(x)T_{\alpha\beta}(z)\rangle++\epsilon_{\sigma\eta\zeta}\partial^{\zeta}\frac{1}{|x-y|^2}\partial_{\rho}\langle T_{\eta\nu}(x)T_{\alpha\beta}(z)\rangle\notag\\
&+\partial_{\nu}\delta^{(3)}(x-z)\langle T_{\sigma\rho}(y)T_{\alpha\beta}(x)\rangle_{odd}+\left\{\partial_{\alpha}(\delta^{(3)}(x-z)\langle T_{\beta\nu}(x)T_{\sigma\rho}(y)\rangle_{odd})+\alpha\leftrightarrow\beta\right\}\notag\\
&\sim\epsilon_{\sigma\eta\zeta}\partial^{\zeta}\partial_{\nu}\frac{1}{|x-y|^2}\langle T_{\eta\rho}(x)T_{\alpha\beta}(z)\rangle + \partial_{\rho}(\epsilon_{\sigma\eta\zeta}\partial^{\zeta}\frac{1}{|x-y|^2}\langle T_{\eta\nu}(x)T_{\alpha\beta}(z)\rangle)\notag\\
&+\epsilon_{\sigma\eta\zeta}\partial^{\zeta}\frac{1}{|x-y|^2}\partial_{\eta}\langle T_{\rho\nu}(x)T_{\alpha\beta}(z)\rangle+\epsilon_{\sigma\eta\zeta}\partial^{\zeta}\frac{1}{|x-y|^2}\partial_{\rho}\langle T_{\eta\nu}(x)T_{\alpha\beta}(z)\rangle
\end{align}
Again, the Ward identity has terms that survive at non-coincident points and therefore, showing that the epsilon transform leads to a non-conserved spin-2 current.

We consider a few more explicit examples of the proposal in \eqref{goutnh1} in the Appendix. \ref{npodapd}. We also discuss converting this new parity odd CFT correlator in spinor helicity variables and connecting to spinor helicity amplitude in the flat-space limit.
To summarize this subsection, we have shown that the new parity odd CFT correlator defined in \eqref{goutnh1} reproduces minimal parity odd flat space amplitude. However, as the table \ref{msmtchcftamp} indicate, even though we have been able to deal with a mismatch of CFT correlator and covariant vertex, the mismatch of counting between CFT correlator and flat space spinor helicity amplitude remains. We summarize this by the following diagram.
\\

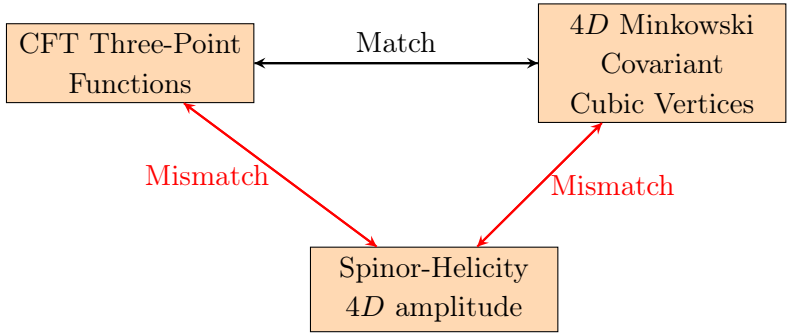
\begin{tikzpicture}[node distance=2cm]
\node (pro1) [process] {CFT Three-Point Functions};
\node (pro2b) [process, right of=pro1, xshift=5cm] {$4D$ Minkowski Covariant\\Cubic Vertices};
\node (pro2c) [process, below of=pro1, xshift=4cm, yshift=-1cm] {Spinor-Helicity $4D$ amplitude};
\draw [arrow] (pro2b) --(pro1);
\draw [arrow] (pro1) -- node[anchor=south] {Match} (pro2b);
\draw [arrow, red] (pro1) -- node[anchor=east] {Mismatch} (pro2c);
\draw [arrow, red] (pro2c) -- (pro1);
\draw [arrow, red] (pro2b) -- node[anchor=west] {Mismatch} (pro2c);
\draw [arrow, red] (pro2c) -- (pro2b);
\end{tikzpicture}

In the next section, we address this issue of mismatch of CFT correlator with spinor helicity variable amplitude.

\section{CFT correlator in spinor helicity/ spinor helicity amplitude correspondence: A closer look at mismatch }\label{rpuza}
The table \ref{msmtchcftamp} highlights the mismatch between spinor helicity amplitude and CFT correlator of conserved currents. This section aims to show that one can construct an extra CFT correlator of conserved currents which in the flat space limit goes over to flat space spinor helicity amplitude.
Interestingly it turns out that these new CFT correlators do not give rise to any consistent covariant 4d amplitude. Let us start the discussion with the simplest of the cases.

\subsection{Two scalar one spinning three point function}
Let us consider CFT correlator of the form $\langle J_{s_1} O O\rangle.$ In the table \ref{msmtchcftamp} it is summarised that there is only one parity even CFT correlator where as amplitude has one parity even and one parity odd component. To take care of this mismatch we define epsilon transformed correlator
\begin{align}\label{prpsl1}
\langle J_{s_1} OO\rangle_{o}=[\mathcal{O}_\epsilon] \langle J_{s_1} OO\rangle_{e}.
\end{align}
As a illustrative example let us consider $\langle T O_2 O_2\rangle$ which is given by
\begin{align}\label{s00cft}
z_3^{\mu}z_3^{\nu}\langle O_2(k_1) O_2(k_2) T_{\mu\nu}(k_3) \rangle_{nh,e}= \frac{ k_1+k_2+2 k_3}{\left(k_1+k_2+k_3\right)^2}\left(k_1.z_3 \right)^2
\end{align}
where $O_2$ is scalar operator with scaling dimension $2$ and $z_3$ is transverse polarization tensor.
The solution in \eqref{s00cft} satisfies ward identity given by
\begin{align}
\langle O O k_3. T \rangle= k_2. z_3 \left( \langle O(k_1) O(-k_1) \rangle - \langle O(k_2) O(-k_2) \rangle \right)
\end{align}
Usign proposal \eqref{prpsl1} we obtain
\begin{align}\label{s00odcft}
z_3^{\mu}z_3^{\nu}\langle O_2(k_1) O_2(k_2) T_{\mu\nu}(k_3) \rangle_{o}= \frac{ k_1+k_2+2 k_3}{k_3\left(k_1+k_2+k_3\right)^2}\left(k_1.z_3 \right)\epsilon(z_3~ k_1~k_3)
\end{align}
In the flat space limit we obtain
\begin{align}\label{s00odcft1}
\lim_{E\rightarrow 0} z_3^{\mu}z_3^{\nu}\langle O_2(k_1) O_2(k_2) T_{\mu\nu}(k_3) \rangle_{o}\sim \frac{ 1}{E^2}\left(k_1.z_3 \right)\epsilon(z_3~ k_1~k_3)+\rm{subleading~ terms}
\end{align} which implies
$A\sim \left(k_1.z_3 \right)\epsilon(z_3~ k_1~k_3)$ which can not be converted to a four dimensional covariant expression \footnote{Four dimensional covariant epsilon will have four free indices. Because of momentum conservation and only one polarization tensor $z_3$ we can't write the expression in terms of four dimensional epsilon tensor.}.
However it is easy to show that in spinor helicity variables \eqref{s00cft}, \eqref{s00odcft} can be written as
\begin{align}\label{s00ab}
\langle O O T_-\rangle &= \left(c_e +i c_o\right) \frac{\left(k_1+k_2+2 k_3\right) \left(k_3 -k_1-k_2\right)^2}{k_3^2\left(k_1+k_2+k_3\right)^2}\left(\frac{\langle 23 \rangle \langle 13 \rangle}{\langle 12 \rangle}\right)^2
\end{align}
and its conjugate. It is clear that in the flat space limit \eqref{s00ab} reproduces both parity even and parity odd amplitude correctly that appears in \eqref{nonpamp}. Again it is easy to show by going to position space that the result in \eqref{s00odcft} corresponds to spin two non conserved current as we shaw in the last section \footnote{More precisely,
the $\langle TOO\rangle$ ward identity in position space is given by
\begin{align}
\partial^{\mu}_1\langle T_{\mu\nu}(x_1)O(x_2)O(x_3)\rangle = \partial_{1\nu}\delta^{(3)}(x_1-x_2)\langle O(x_1)O(x_3)\rangle+\partial_{1\nu}\delta^{(3)}(x_1-x_3)\langle O(x_1)O(x_2)\rangle
\end{align}
The epsilon transform of the above yields
\begin{align}
[O_{\epsilon}]_1 \partial^{\mu}_1\langle T_{\mu\nu}(x_1)O(x_2)O(x_3)\rangle &=\epsilon_{\nu\alpha\tau}\left(\partial_{1\tau}\frac{1}{|x_{12}|^2}\partial_{2\alpha}\langle O(x_2) O(x_3)\rangle+\partial_{1\tau}\frac{1}{|x_{13}|^2}\partial_{3\alpha}\langle O(x_3) O(x_2)\rangle\right)
\end{align}
which is non-zero even away from insertion of the operators. }.

It turns out that story in CFT side is much more richer. One can also find more solution to conformal ward identity \eqref{cwt1} for $\langle OO J_s\rangle.$ Solution \eqref{s00cft} is non-homogeneous, one can also find homogeneous solutions
\begin{align}\label{OOThs}
\langle O O T_-\rangle =\left( c'_e+ i c'_o\right)\frac{k_3}{ E^2} \left(\frac{\langle 23 \rangle \langle 13 \rangle}{\langle 12 \rangle}\right)^2.
\end{align}
If we convert the solution \eqref{OOThs} in momentum space we obtain
\begin{align}\label{ottbdpol}
\langle O O T\rangle= \frac{k_3^3}{\left(k_1+k_2+k_3\right)^2 \left(k_3-k_1-k_2\right)^2 } \left(z_3.k_1\right)^2
\end{align}
which has bad pole at $k_3=k_1+k_2$ and is not consistent
with position space OPE limit \cite{Maldacena:2011nz,Jain:2021whr}. However, the flat space limit produces correct spinor helicity amplitude.

We now consider another example $\langle JJ J_4\rangle.$ In this case, we have considered correlators that saturate the WT identity, see Appendix \ref{npodapd} for details. To construct a new CFT correlator we need to construct a homogeneous solution that does not contribute to WT identity as we did above.

\subsection{General spin $s_1\neq s_2\neq s_3\neq 0$}
It can be shown that for a general spin as well the flat space limit, in general, reproduces the correct covariant vertex but not the full amplitude in spinor helicity variables. This implies the flat space limit of the CFT correlator can not reproduce all the spinor helicity amplitudes. As is discussed earlier, in this case, we have four CFT correlators, three coming from conserved currents and one constructed in \eqref{goutnh1}. However, spinor helicity amplitude has a total of eight independent structures. It is easy to establish that the mismatch is in the helicity components $+-+, -+-$ and in $--+,++-$ as is the case in \eqref{cubv1} and \eqref{cubva}. In the following, we show that this can be remedied by considering homogeneous solutions \footnote{Let us note that, we can't have more non-homogeneous solutions as we have already saturated the WT identity with non-homogeneous parity even solution.}.

Consider the following ansatz for $\langle J_{s_1} J_{s_2} J_{s_3}\rangle$ in spinor-helicity variables \cite{Jain:2021whr}
\begin{align}
\left\langle J_{s_{1}}^{h_{1}} J_{s_{2}}^{h_{2}} J_{s_{3}}^{h_{3}}\right\rangle=f_{h_{1}, h_{2}, h_{3}}\left(k_{1}, k_{2}, k_{3}\right)\langle 12\rangle^{h_{3}s_3-h_{1}s_1-h_{2}s_2}\langle 23\rangle^{h_{1}s_1-h_{2}s_2-h_{3}s_3}\langle 31\rangle^{h_{2}s_2-h_{3}s_3-h_{1}s_1}
\end{align}
Using the homogeneous ward identity, one finds that the most general solution to $f$ are of two kinds and is given by \cite{Jain:2021whr}
\begin{align}\label{solarbt}
&f^1_{h_{1}, h_{2}, h_{3}}(k_1, k_2, k_3) = \frac{k^{s_1-1}_1k^{s_2-1}_2k^{s_3-1}_3}{E^{-h_1s_1-h_2s_2-h_3s_3}}\nonumber\\
&f^2_{h_{1}, h_{2}, h_{3}}(k_1, k_2, k_3) = \frac{k^{s_1-1}_1k^{s_2-1}_2k^{s_3-1}_3}{(k_1+k_2-k_3)^{h_{3}s_3-h_{1}s_1-h_{2}s_2}(k_3+k_2-k_1)^{h_{1}s_1-h_{2}s_2-h_{3}s_3}(k_1+k_3-k_2)^{h_{2}s_2-h_{1}s_1-h_{3}s_3}}.
\end{align}
Except in all $-$ or all $+$ helicity, both the solutions when converted to momentum space  in \eqref{solarbt} have bad poles and are not consistent with the position space OPE limit, see \cite{Jain:2021whr} for more discussion \footnote{It is important to note that if we discard consistency with OPE, CFT allows for more structures in spinor helicity variables than the flat space spinor helcity amplitude. However,  the solutions $f^2$ has explicit bad poles even before converting to momentum space variables.}. For all $-$ or all $+$ helicity also only the first solution in \eqref{solarbt} is consistent. We can use the solution in \eqref{solarbt} in the flat space limit to reproduce missing helicity components $+-+, -+-$ and $--+,++-.$ However as mentioned, it can be shown easily that they don't reproduce any local covariant vertex. Also, the CFT correlator corresponding to these helicities are not consistent with the position space OPE limit because of the presence of bad poles \cite{Jain:2021whr}.

Notice that in the flat-space limit we have
\begin{align}\label{cftfltspa}
\lim_{E\to 0} \left\langle J_{s_{1}}^{h_{1}} J_{s_{2}}^{h_{2}} J_{s_{3}}^{h_{3}}\right\rangle\sim\langle 12\rangle^{h_{3}s_3-h_{1}s_1-h_{2}s_2}\langle 23\rangle^{h_{1}s_1-h_{2}s_2-h_{3}s_3}\langle 31\rangle^{h_{2}s_2-h_{3}s_3-h_{1}s_1}
\end{align}
we obtain the non-perturbative amplitude. See Appendix \ref{fltsp1} for more details of flat space limit of CFT correlators in spinor helicity variables.
The above result implies that the CFT correlator has more solutions and in the flat space limit it reproduces all possible spinor helicity amplitudes. The following table summarizes the counting.

\begin{table}[h!]
\begin{tabular}{||c | c | c ||}
\hline
\multicolumn{3}{|c|}{CFT correlator vs spinor helicity amplitude}\\
\hline\hline
$\mathbf{s_1 \leq s_2 \leq s_3\neq 0}$ & \begin{tabular}{@{}c@{}}\bf{CFT correlator}\\\bf{of conserved currents} \end{tabular} & \begin{tabular}{@{}c@{}}\bf{Spinor Helicity}\\\bf{Amplitude} \end{tabular} \\ [0.5ex]
\hline
$s_1 \ne s_2 \ne s_3$ & 4 h even+ 4 h odd+ 1 nh even + 1 nh odd & 4 even , 4 odd\\
\hline
\end{tabular}
\caption{This table summarizes CFT correlator counting and flat space spinor helicity amplitude in most general cases. Let us note that we have ignored the second homogeneous solution in \eqref{solarbt}. We have also ignored non-homogeneous parity odd contact terms analogous to \eqref{contnhod}. In the table h stands for homogeneous solution and nh stands for non-homogeneous solution. Except for the nh odd solution in this table, all the other solutions are for conserved currents. The nh odd solution is defined in \eqref{goutnh1}. }
\label{T6a}
\end{table}
We also work out an explicit example Appendix \ref{411ab}.

\bigskip

\section{Connection to AdS amplitude}\label{adsfltab}
Three-point amplitudes in AdS was calculated in spinor helicity variables in \cite{Nagaraj:2018nxq,Nagaraj:2019zmk} and in light cone variables gauge in \cite{Metsaev:2018xip}. This section aims to make a connection with CFT results presented in previous sections with results in AdS \footnote{We thank D. Ponomarev, E. Skvortsov for their insightful comments which led to better understanding of the content of this section.}. One of the main issues with identification is it is not clear which amplitude in AdS is identified with a homogeneous/ non-homogeneous CFT correlator.
One way to get around this difficulty is to make use of the connection between AdS amplitude and flat space amplitude. Since we also know which flat space amplitude corresponds to which CFT correlator, this gives us an indirect way to connect the CFT correlator to AdS amplitudes.
Let us take an example to illustrate this connection.
In \cite{Nagaraj:2018nxq,Nagaraj:2019zmk} a simple relation between flat space and AdS amplitude in spinor helicity variables was pointed out. For example for the case of $spin-0-0-2$ the explicit relation is given in eq.(7.23) of \cite{Nagaraj:2019zmk}
\begin{align}\label{fladsc}
A_{-}\sim \left(\frac{\langle 23 \rangle \langle 13 \rangle}{\langle 12 \rangle}\right)^2 \left(1+ \frac{\Box_{k}}{R_{AdS}^2}\right) \delta^{4}(k).
\end{align}
which $A_-$ stands for amplitude in minus helicity.
It is clear that in the limit $R_{AdS} \rightarrow \infty$ we get the flat space amplitude\footnote{ Let us note that corresponding CFT correlator is as given in \eqref{s00ab}, \eqref{OOThs}. It is not clear if the result in \eqref{fladsc} corresponds to a non-homogenous solution in \eqref{s00ab} or homogeneous solution in \eqref{OOThs}. Let us note that \eqref{fladsc} can be generalized for other correlators and in general can be stated as
\begin{align}\label{fladsc1}
A_{AdS}\sim A_{flat~space}\left(1+ \frac{\Box_{k}}{R_{AdS}^2}\right) \delta^{4}(k).
\end{align}
For general spin, the connection between the CFT correlator and AdS is more clear in the helicity component $+-+,-+-$ and it's conjugate as for CFT correlator, only homogeneous solution that exists in this case.
}. Let us note that the corresponding CFT correlator is as given in \eqref{s00ab}, \eqref{OOThs}.

Amplitudes in AdS in light-cone was calculated in \cite{Metsaev:2018xip}. In \cite{Skvortsov:2018uru} AdS results were related to CFT amplitudes. Below we describe briefly the results that are important for us. Consider the $AdS_4$ metric in the light-cone coordinates
\begin{align}
ds^2 = \frac{2dx^{+}dx^{-}+ dz^2+dx^2_1}{z^2}
\end{align}
where
\begin{align}
x^{\pm} = \frac{x_2\pm x_0}{\sqrt{2}}
\end{align}
For a massless arbitrary spin-$s$ field, the fourier transform implemented in these coordinates is \cite{Metsaev:2018xip}
\begin{align}
\phi_{\lambda}(x^{+}, x^{-}, x_1, z) = \int \frac{dk_1 d\beta}{2\pi} e^{i(k_1 x_1 + \beta x^{-})}\phi_{\lambda}(x^{+}, \beta, k_1, z), \quad \lambda = \pm s\label{ft}
\end{align}
The fields also satisfy
\begin{align}
&\phi_{s}(x^{+}, x^{-}, x_1, z)^{\dagger} = \phi_{-s}(x^{+}, x^{-}, x_1, z)
\end{align}
The most general cubic vertices made out of $\phi_{\lambda}(x^{+}, \beta, k_1, z)$ is given by
\begin{align}
V^{\lambda_{1}, \lambda_{2}, \lambda_{3}}= \begin{cases}
&V^{\lambda_1, \lambda_2, \lambda_3}_R, ~~H>0 \\
& V^{\lambda_1, \lambda_2, \lambda_3}_L, ~~H<0
\end{cases}
\end{align}
where $H=h_1 s_1+ h_2 s_2+ h_3 s_3.$
We refer interested resders to eq.(1.19) of \cite{Skvortsov:2018uru} for detailed discussions. In the flat-space limit, this leads to
\begin{align}\label{ll}
&\left.V_{R}^{\lambda_{1}, \lambda_{2}, \lambda_{3}}\right|_{\text {flat limit }} \sim[12]^{\lambda_{1}+\lambda_{2}-\lambda_{3}}[23]^{\lambda_{2}+\lambda_{3}-\lambda_{1}}[13]^{\lambda_{1}+\lambda_{3}-\lambda_{2}}, & H>0, \nonumber\\
&\left.V_{L}^{\lambda_{1}, \lambda_{2}, \lambda_{3}}\right|_{\text {flat limit }} \sim\langle 12\rangle^{-\lambda_{1}-\lambda_{2}+\lambda_{3}}\langle 23\rangle^{-\lambda_{2}-\lambda_{3}+\lambda_{1}}\langle 13\rangle^{-\lambda_{1}-\lambda_{3}+\lambda_{2}}, & H<0,
\end{align}
which is equivalent to the statement made in \eqref{cftfltspa} where we took the flat space limit of the CFT correlator to get the flat space helicity amplitude.
The AdS vertex can formally be related to CFT correlator by following identification \cite{Skvortsov:2018uru}
\begin{align}\label{3ptadsa}
\int_{AdS_4} V^{\lambda_1, \lambda_2, \lambda_3} =\sum_{n=0}^{|H|-1} \mathcal{Q}_{n}^{H} \Gamma(|H|-n)\left(\frac{1}{3 \sqrt{2} E}\left(\sum_{a} \breve{\beta}_{a}\left(\left|k_{a}\right| \pm k_{a}^{1}\right)\right)\right)^{|H|-n} \sim \langle\mathbb{J}_{\lambda_1}\mathbb{J}_{\lambda_2}\mathbb{J}_{\lambda_3}\rangle
\end{align}
see eq.(2.15) of \cite{Skvortsov:2018uru} for details. What is important here is appearance of the pole in $\frac{1}{E}.$ There are 8 different structures in the above equation\footnote{Namely, $(\lambda_1, \lambda_2, \lambda_3) = \{---, --+, -+-, +--\}$ and their conjugates. Each helicity structure and their conjugate gives rise to a parity even and a parity odd amplitude.}. This equation is very similar to the solution obtained in (\ref{solarbt}). Even though there is a $1/E$ pole in the result \eqref{3ptadsa}, when we convert this to covariant momentum space expressions as in previous section, it can have bad poles as we saw in (\ref{ottbdpol})\footnote{The relation in \eqref{3ptadsa} is very formal and needs to be understood more properly. For example, it is not very clear how to use the relation in \eqref{3ptadsa} to obtain what we define various homogeneous or non-homogeneous CFT correlators.} so they in general are not consistent with position space OPE limit. Some particular combination of the correlator will correspond to acceptable homogeneous and non-homogeneous results \footnote{Interstingly, study of mismatch between covariant vertex and spinor helicity/light cone amplitudeIn was explored also in $AdS_4$ and it eventually lead to understanding of Chiral HS theory \cite{Skvortsov:2018jea,Krasnov:2021nsq}. It would be interesting to understand analougous statement in CFT side.
}.

For example, in \cite{Skvortsov:2018uru} spin-2 case $\langle TTT\rangle $ was discussed explicitly. For CFT, $\langle TTT\rangle$ results are presented in spinor helicity variables in \eqref{tttsph1}. As is clear the homogeneous part appears only in helicity $---, +++$ where as non-homogeneous piece has non-trivial contribution also in mixed helicity such as $--+,++-.$ The homogeneous parity even and odd parts arises due to weyl tensor cubed $W^3$ and from ${\tilde W}W^2$ respectively where as non-homogeneous parity even contribution is given by Einstein term $\sqrt{g}R$ from $AdS$ or $dS$ perspective. It was shown in \cite{Skvortsov:2018uru} that $V^{+2,-2,-2}+ V^{-2,+2,+2}$ corresponds to non-homogeneous or Einstein gravity term where as $V^{-2,-2,-2}\pm V^{+2,+2,+2}$ contributes to parity even and parity odd homogeneous term or Weyl cubed term. This is consistent with results in spinor helicity variables in \eqref{tttsph1}. Using this correspondence, one can in general for $\langle J_{s_1} J_{s_2} J_{s_3} \rangle$ one can argue that \footnote{At least when spin satisfy triangle inequaliuty $s_i\le s_j+s_k$. For correlator violating triangle inequality, things are little more complicated. See \cite{Jain:2021whr} for more details.} all negative or all positive helicity components in \eqref{3ptadsa} will correspond to parity even and parity odd acceptable homogeneous solution. Some of the mixed helicity components will lead to non-homogenous acceptable CFT correlator and other mixed helicity component will lead to homogeneous CFT correlator with bad poles. For specific cases this relation can be made more precise.

To conclude, we observe that relation in \eqref{3ptadsa} gives rise to in general known CFT correlators which are consistent with position space OPE limit as well as some other CFT correlator which contains bad poles and hence are not consistent with position space OPE limit.

\section{Summary and Discussion}\label{disc}
This paper is devoted towards understanding the CFT correlation function, scattering amplitude connection. In particular we focused on connection between 3 point CFT correaltion function of conserved currents with flat space three point amplitude of mass less gauge fields.  In particular we show the following map
\begin{align}\label{mapamp11}
\langle J_{s_1} J_{s_2}J_{s_3}\rangle_{nh,e}&\rightarrow M^{s_1 s_2 s_3}_{m,e}\nonumber\\
\langle J_{s_1} J_{s_2}J_{s_3}\rangle_{h,e}&\rightarrow M^{s_1 s_2 s_3}_{nm,e}\nonumber\\
\langle J_{s_1} J_{s_2}J_{s_3}\rangle_{h,o}&\rightarrow M^{s_1 s_2 s_3}_{nm,o}.
\end{align}
This map indicates that number of CFT correlation function of conserved currents are always less than 
the number of allowed structures for flat space amplitude. In spinor helcity variables this mismatch is even more.  Table \ref{msmtchcftamp} summarises the known results. We also have discussed a similar issue with $AdS_4$ amplitude in spinor helicity variables and light cone gauge. We then point out parity odd CFT structure which in the flat space limit goes over to missing flat space covariant vertex. However we showed that this extra parity odd CFT correlator can not be constructed out of conserved currents as expected. Interestingly, this extra CFT correlator is consistent with the position space OPE limit. We then addressed the issue of mismatch of CFT correlator in spinor helicity variables with amplitude in spinor helicity variables.
We constructed extra CFT correlators which in the flat space limit gives rise to correct spinor helicity amplitudes but do not give rise to any new covariant vertex. We showed these extra CFT structures are not consistent with the position space OPE limit.
We also briefly discussed connection of newly constructed CFT structures with amplitude in $AdS_4$ in light cone gauge as well as in spinor helicity variables.

One can also establish a double copy relation involving a new parity-odd non-homogeneous CFT correlator that we defined using epsilon transformation. Let us briefly discuss this here.
\section*{Double copy relation}
For the homogeneous piece we have in spinor helicity variables
\begin{equation}\label{hevod}
\langle J_{s_1} J_{s_2}J_{s_3}\rangle_{\bf{h}, e}\propto \langle J_{s_1} J_{s_2}J_{s_3}\rangle_{\bf{h}, o}.
\end{equation}
Using such properties it is easy to establish in momentum space
\cite{Jain:2021qcl}
\begin{align}
\langle J_{2s_1} J_{2s_2}J_{2s_3}\rangle_{\bf{h}, e}&\propto \left( \langle J_{s_1} J_{s_2}J_{s_3}\rangle_{even, \bf{h}}\right)^2 \propto \left(\langle J_{s_1} J_{s_2}J_{s_3}\rangle_{\bf{h}, o}\right)^2\nonumber\\
\langle J_{2s_1} J_{2s_2}J_{2s_3}\rangle_{\bf{h}, o} &\propto \langle J_{s_1} J_{s_2}J_{s_3}\rangle_{\bf{h}, e} \langle J_{s_1} J_{s_2}J_{s_3}\rangle_{ \bf{h}, o}
\end{align}
using which we get
\begin{align}
\langle J_{2s_1} J_{2s_2}J_{2s_3}\rangle_{\bf{h}, o}+\langle J_{2s_1} J_{2s_2}J_{2s_3}\rangle_{o, \bf{h}}&\propto \left( \langle J_{s_1} J_{s_2}J_{s_3}\rangle_{\bf{h}, e}+\langle J_{s_1} J_{s_2}J_{s_3}\rangle_{\bf{h}, o}\right)^2.
\end{align}
It was also shown that
non-homogeneous structures separately satisfy double copy relation
\begin{align}
\langle J_{2s_1} J_{2s_2}J_{2s_3}\rangle_{ \bf{nh}, e}&\propto \left( \langle J_{s_1} J_{s_2}J_{s_3}\rangle_{\bf{nh}, e}\right)^2
\end{align}
As discussed in the main text, even for non-homogeneous parity-odd and even pieces defined in \eqref{goutnh1} we have analogue of \eqref{hevod} in spinor helicity variables
\begin{equation}\label{hevod1}
\langle J_{s_1} J_{s_2}J_{s_3}\rangle_{\bf{nh}, e}\propto \langle J_{s_1} J_{s_2}J_{s_3}\rangle_{\bf{nh}, o}.
\end{equation}
This again implies
\cite{Jain:2021qcl}
\begin{align}
\langle J_{2s_1} J_{2s_2}J_{2s_3}\rangle_{\bf{nh}, e}&\propto \left( \langle J_{s_1} J_{s_2}J_{s_3}\rangle_{\bf{nh}, e}\right)^2 \propto \left(\langle J_{s_1} J_{s_2}J_{s_3}\rangle_{\bf{nh}, o}\right)^2\nonumber\\
\langle J_{2s_1} J_{2s_2}J_{2s_3}\rangle_{\bf{nh}, o} &\propto \langle J_{s_1} J_{s_2}J_{s_3}\rangle_{\bf{nh}, e} \langle J_{s_1} J_{s_2}J_{s_3}\rangle_{\bf{nh}, o}
\end{align}
using which we get
\begin{align}
\langle J_{2s_1} J_{2s_2}J_{2s_3}\rangle_{\bf{nh}, e}+\langle J_{2s_1} J_{2s_2}J_{2s_3}\rangle_{\bf{nh}, o}&\propto \left( \langle J_{s_1} J_{s_2}J_{s_3}\rangle_{\bf{nh}, e}+\langle J_{s_1} J_{s_2}J_{s_3}\rangle_{\bf{nh}, o}\right)^2.
\end{align}

It would be interesting to generalize the discussion for four point functions. However, this will be a significantly harder problem to address. In \cite{Jain:2020rmw,Jain:2020puw} using higher spin equations in momentum space, CFT three-point functions of higher spin operators and four point functions of few simple cases were explored. It is easy to show that the three point covariant vertex, the minimal even, non-minimal even and non-minimal odd amplitudes satisfy the same higher spin equations. Infact, one uses these higher spin equations to calculate them. It will be interesting to use this strategy to calculate four point covarint amplitudes of higher spin fields as well as four point CFT correlators of higher spin operators. We should also be able to check easily if the epsilon transform that we have discussed in this paper can be used at the level of amplitudes which has been classified recently in \cite{Chowdhury:2019kaq}. In general the study of flat space amplitude/dS or AdS amplitude/ CFT correlator correspondence \footnote{See \cite{Rastelli:2016nze,Arkani-Hamed:2017fdk,Armstrong:2020woi,Albayrak:2020fyp,Gillioz:2020mdd,Eberhardt:2020ewh,Roehrig:2020kck,Zhou:2021gnu,Alday:2022lkk,Alday:2021odx,Herderschee:2022ntr,Cheung:2022pdk} and references therein for some recent development.} is an important one has led to important developments and there are many important avenues open to be explored.

\section*{Acknowledgments}
The work of S.J is supported by the Ramanujan Fellowship.
AM would like to acknowledge the support of CSIR-UGC
(JRF) fellowship (09/936(0212)/2019-EMR-I).
We thank S.C-Hout, E. Joung, K. Mkrtchyan for valuable email exchanges explaining their earlier works and comments on a previous version. We specially thank D. Ponomarev, E. Skvortsov for extensive email exchange explaining their works, related issues and extensive comments on a previous version of the draft. We would like to thank S. Ananth for the valuable discussion.
We acknowledge our debt to the people of India for their steady support of research in basic sciences.
	
	\appendix

\section{Flat-space amplitudes: Examples}\label{ampflt}
In this section, we give some simple examples of flat-space $4D$ scattering amplitude. We define two sets of amplitudes, one that satisfies $s_i\le s_{j}+s_{k}$ is called inside the triangle and one that violates this is called outside the triangle. This distinction becomes very important for momentum space CFT correlators as was discussed in \cite{Jain:2021whr}.

\subsection{Inside the triangle inequality}
We take a simple example of two photon and graviton scattering. The results are given by four structures, two parity-even and one parity-odd
\begin{align}\label{evamp1}
\mathcal{M}^{112}_{even} = &g_{m, e} (z_1.p_2 z_2.z_3+z_2.p_3 z_3.z_1+z_3.p_1z_1.z_2)(z_3.p_1)+g_{nm, e} (z_1.p_2)(z_2.p_3)(z_3.p_1)^2\\
\mathcal{M}^{112}_{odd} = &g_{m, o}[\epsilon(z_2p_2z_3p_3)(z_3.z_1)+\epsilon(z_3p_3z_1p_1)(z_2.z_3)+(\epsilon(z_1z_2z_3p_2)-\epsilon(z_1z_2z_3p_1))(z_3.k_1)]\notag\\& +g_{nm, o}\epsilon(z_2p_2z_3p_3)(z_1.p_2)(z_3.p_1)
\end{align}
Both the minimal and non-minimal amplitudes are present for both parity-even and parity-odd cases. Notice that the odd minimal amplitude is antisymmetric under $1 \leftrightarrow 2$ exchange. Therefore, one needs to introduce Chan-Paton factors for that amplitude. In fact, it turns out Chan-Paton factors must be introduced for amplitudes with $s_1 = s_2 < s_3$ for even $s_3$. We rewrite the odd amplitude in $3D$ momentum space variables \eqref{3d2d}
\begin{align}
\mathcal{M}^{112}_{odd} = &g_{m, o}[-\left(\epsilon\left(z_{2} z_{3} k_{2}\right) k_{3}-\epsilon\left(z_{2} z_{3} k_{3}\right) k_{2}\right)\left(z_{1} \cdot z_{3}\right)+\epsilon\left(z_{1} z_{2} z_{3}\right) k_{2}\left(z_{3} \cdot k_{1}\right)]\notag\\& +g_{nm, o}[-\left(\epsilon\left(z_{2} z_{3} k_{2}\right) k_{3}-\epsilon\left(z_{2} z_{3} k_{3}\right) k_{2}\right)](z_1.k_2)(z_3.k_1)\label{JJT3D}
\end{align}

It is easy to show under epsilon transform that the parity-even amplitude in \eqref{evamp1} maps to parity-odd amplitude \eqref{JJT3D}. Let us for completeness show this below explicitly.
Consider
\begin{align}
[O_{\epsilon}]_2 \mathcal{M}^{112}_{e} &= g_{m, e}(z_1.k_2 \frac{\epsilon(z_2k_2z_3)}{k_2}+z_3.z_1 \frac{\epsilon(z_2k_2k_3)}{k_2}+z_3.k_1\frac{\epsilon(z_2 k_2 z_1)}{k_2})(z_3.k_1)\notag\\&+g_{nm, e} (z_1.k_2)\frac{\epsilon(z_2 k_2 k_3)}{k_2}(z_3.k_1)^2
\end{align}
Using the Schouten identities
\begin{align}
&(z_1.k_2) \epsilon(z_2 k_2 z_3) = k^2_2\epsilon(z_2 z_1 z_3)-z_3.k_1\epsilon(z_2 k_2 z_1)\\
&(z_3.k_1) \epsilon(z_2 k_2 k_3) = -k^2_2\epsilon(z_2 z_3 k_3)-k_2 k_3 \epsilon(z_2 k_2 z_3)
\end{align}
in the above we exactly get (\ref{JJT3D}). We have also made use of $ k_I k_J=k_I.k_J $ in the above. A more abstract derivation of the same is given in Section \ref{eptr}. From here on-wards, we write all the flat-space amplitudes in $3D$ momentum space variables.
\subsection*{Three spin-s amplitude}
For general three spin-s amplitude we have
\begin{align}
\mathcal{M}^{sss}_{e} &= g_{m, e} (z_1.k_2 z_2.z_3+z_2.k_3 z_3.z_1+z_3.k_1z_1.z_2)^s+g_{nm, e} (z_1.k_2)^s(z_2.k_3)^s(z_3.k_1)^s\nonumber\\
\mathcal{M}^{sss}_{o} &= g_{nm, o}[-\epsilon(z_2 z_3 k_2) k_3+\epsilon(z_2 z_3 k_3) k_2](z_1.k_2)^s(z_2.k_1)^{s-1}(z_3.k_1)^{s-1}
\end{align}
Notice no minimal amplitude present for the parity-odd case. This is as mentioned before in Table \ref{Tamp2a}. The minimal term was dropped as the negative powers appearing due to $s_1 = s_2 = s_3.$
For example, three photon amplitude is given by
\begin{align}
\mathcal{M}^{111}_{e} &= g_{m, e} (z_1.k_2 z_2.z_3+z_2.k_3 z_3.z_1+z_3.k_1z_1.z_2)+g_{nm, e} (z_1.k_2)(z_2.k_3)(z_3.k_1)\nonumber\\
\mathcal{M}^{111}_{o} &= g_{nm, o}[-\epsilon(z_2 z_3 k_2) k_3+\epsilon(z_2 z_3 k_3) k_2](z_1.k_2)
\end{align}
Another simple example is of three graviton scattering where we have
\begin{align}
\mathcal{M}^{222}_{e} &= g_{m, e} (z_1.k_2 z_2.z_3+z_2.k_3 z_3.z_1+z_3.k_1z_1.z_2)^2+g_{nm, e} (z_1.k_2)^2(z_2.k_3)^2(z_3.k_1)^2\nonumber\\
\mathcal{M}^{222}_{o} &= g_{nm, o}[-\epsilon(z_2 z_3 k_2) k_3+\epsilon(z_2 z_3 k_3) k_2](z_1.k_2)^2(z_2.k_1)(z_3.k_1)
\end{align}

\subsubsection*{One scalar two spin-s amplitude}
For two photon and one scalar we have only two structures
\begin{align}\label{110}
\mathcal{M}^{011}_{e} = g_{nm, e} (z_2.k_3)(z_3.k_1),~~
\mathcal{M}^{011}_{o} = g_{nm, o}[-\epsilon(z_2 z_3 k_2) k_3+\epsilon(z_2 z_3 k_3) k_2]
\end{align}
this can be generalised to two spin-s and one scalar.
\begin{align}
\mathcal{M}^{0ss}_{e} = g_{nm, e} (z_2.k_3)^s(z_3.k_1)^s,~~
\mathcal{M}^{0ss}_{o} = g_{nm, o}[-\epsilon(z_2 z_3 k_2) k_3+\epsilon(z_2 z_3 k_3) k_2](z_2.z_3)^{s-1}
\end{align}

\subsection{Outside the triangle inequality}
Let us consider one spin-s two scalar amplitude
\begin{align}
\mathcal{M}^{00s}_{e} = g_{e}(z_3.k_1)^s.
\end{align}
This only has parity-even contribution. For one scalar one spin $s_2$ and one spin $s_3$ amplitude we have
\begin{align}
\mathcal{M}^{0s_2s_3}_{e} =g_{e} (z_2.k_1)^{s_2}(z_3.k_1)^{s_3},~~~
\mathcal{M}^{0s_2s_3}_{o} = g_{o} [-\epsilon(z_2 z_3 k_2) k_3+\epsilon(z_2 z_3 k_3) k_2] (z_2.k_1)^{s_2-1}(z_3.k_1)^{s_3-1}
\end{align}
with $s_3 > s_2.$
As an example let us consider spin-3 spin-1 scalar amplitude which is given by
\begin{align}\label{310}
\mathcal{M}^{013}_{e} = g_{e} (z_2.k_1)(z_3.k_1)^3,~~~~
\mathcal{M}^{013}_{o} = g_{o} [-\epsilon(z_2 z_3 k_2) k_3+\epsilon(z_2 z_3 k_3) k_2] (z_3.k_1)^2.
\end{align}

Now let us consider two spin-1 and one spin-4 particles which will be useful for our purposes. We have
\begin{align}\label{114ampa}
\mathcal{M}^{114}_{e} &= g_{m, e} (z_1.k_2 z_2.z_3+z_2.k_3 z_3.z_1+z_3.k_1z_1.z_2)(z_3.k_1)^3+g_{nm, e} (z_1.k_2)(z_2.k_3)(z_3.k_1)^3\nonumber\\
\mathcal{M}^{114}_{o} = &g_{m, o}[-\left(\epsilon\left(z_{2} z_{3} k_{2}\right) k_{3}-\epsilon\left(z_{2} z_{3} k_{3}\right) k_{2}\right)\left(z_{1} \cdot z_{3}\right)+\epsilon\left(z_{1} z_{2} z_{3}\right) k_{2}\left(z_{3} \cdot k_{1}\right)](z_3.k_1)^2\notag\\& +g_{nm, o}[-\epsilon(z_2 z_3 k_2) k_3+\epsilon(z_2 z_3 k_3) k_2](z_1.k_2)(z_3.k_1)^3.
\end{align}

\section{Flat-space limits of CFT correlators}\label{CFTfta}
In this section, we shall discuss with some examples how the flat-space amplitude can be obtained from the CFT correlator in momentum space.
We again break up the discussion into inside the triangle and outside the triangle. The details of the CFT correlators can be found in \cite{Jain:2021vrv}.
\subsection{Within triangle inequality}
Let us start with the simplest of examples.
\subsubsection*{$\langle O J_s J_s\rangle$}
In momentum space, we have
\begin{align}
&\langle J_sJ_s O_{\Delta}\rangle_{e} =\left(k_{1} k_{2}\right)^{s-1} I_{\frac{1}{2}+2 s\left\{\frac{1}{2}, \frac{1}{2}, \Delta-\frac{3}{2}\right\}}\left[2\left(\vec{z}_{1} \cdot \vec{k}_{2}\right)\left(\vec{z}_{2} \cdot \vec{k}_{1}\right)+E\left(E-2 k_{3}\right) \vec{z}_{1} \cdot \vec{z}_{2}\right]^{s}\\
&\langle J_sJ_s O_{\Delta}\rangle_{o} =\left(k_{1} k_{2}\right)^{s-1} I_{\frac{1}{2}+2 s\left\{\frac{1}{2}, \frac{1}{2}, \Delta-\frac{3}{2}\right\}}\left[k_{2} \epsilon^{k_{1} z_{1} z_{2}}-k_{1} \epsilon^{k_{2} z_{1} z_{2}}\right] \notag\\
&\times\left[2\left(\vec{z}_{1} \cdot \vec{k}_{2}\right)\left(\vec{z}_{2} \cdot \vec{k}_{1}\right)+E\left(E-2 k_{3}\right) \vec{z}_{1} \cdot \vec{z}_{2}\right]^{s-1}
\end{align}
which in the flat-space limit $E \to 0$ leads to
\begin{align}
&\lim_{E \to 0}\langle J_sJ_sO_{\Delta}\rangle_{e} \sim\frac{\left(k_{1} k_{2}\right)^{s-1}k^{\Delta-2}_3}{E^{2s}} (z_1.k_2 z_2.k_1)^{s}\\
&\lim_{E \to 0}\langle J_sJ_sO_{\Delta}\rangle_{o} \sim\frac{\left(k_{1} k_{2}\right)^{s-1}k^{\Delta-2}_3}{E^{2s}} [\epsilon(z_1z_2k_1)k_2-\epsilon(z_1z_2k_2)k_1](z_1.k_2 z_2.k_1)^{s-1}
\end{align}
where we can immediately identify $\mathcal{A}^{ss0}_{even}$ and $\mathcal{A}^{ss0}_{odd}$ amplitudes where in the last line we have used \eqref{3d2d} to convert the amplitude to four dimensional language.

\begin{table}[h!]
\begin{tabular}{||c | c | c ||}
\hline
\multicolumn{3}{|c|}{Inside the triangle $\langle O J_{s}J_{s}\rangle$ (Exactly conserved current)}\\
\hline\hline
\begin{tabular}{@{}c@{}}\bf{$4D$ Flat-space} \\ \bf{Amplitudes}\end{tabular} & \bf{3D CFT correlator} & \begin{tabular}{@{}c@{}}\bf{Expected CFT} \\ \bf{Pole structure}\end{tabular} \\ [0.5ex]
\hline
\begin{tabular}{@{}c@{}} Even\end{tabular} & Free Boson & $E^{-2s}$ \\
\hline
\begin{tabular}{@{}c@{}} Odd\end{tabular}& Free Fermion & $E^{-2s}$ \\ [1ex]
\hline
\end{tabular}
\caption{$\langle O J_{s}J_{s}\rangle$ has only Homogeneous component and the flat-space amplitude can be obtained from free theories.}
\label{T21}
\end{table}

\subsubsection*{$\langle J_s J_s J_s\rangle$}
This correlator has homogeneous and non-homogeneous contribution \cite{Jain:2021vrv}. Homogeneous piece has both parity-even and parity-odd contribution whereas non-homogeneous piece has only parity-even contribution. The parity-even homogeneous and non-homogeneous piece can be obtained from
\begin{align}\label{hnbf}
\langle J_s J_s J_s\rangle_{e,h}&= \langle J_s J_s J_s\rangle_{e,FB} -\langle J_s J_s J_s\rangle_{e,FB} \nonumber\\
\langle J_s J_s J_s\rangle_{e,h}&= \langle J_s J_s J_s\rangle_{e,FB} +\langle J_s J_s J_s\rangle_{e,FB}
\end{align}
The homogeneous piece answer can be written down as follows
\begin{align}
&\left\langle J_{s} J_{s} J_{s}\right\rangle_{\text {e }, \mathbf{h}}=\left(k_{1} k_{2} k_{3}\right)^{s-1}\left[\frac{1}{E^{3}}\left\{2\left(\vec{z}_{1} \cdot \vec{k}_{2}\right)\left(\vec{z}_{2} \cdot \vec{k}_{3}\right)\left(\vec{z}_{3} \cdot \vec{k}_{1}\right)+E\left\{k_{3}\left(\vec{z}_{1} \cdot \vec{z}_{2}\right)\left(\vec{z}_{3} \cdot \vec{k}_{1}\right)+\text { cyclic }\right\}\right\}\right]^{s}\\
&\left\langle J_{s} J_{s} J_{s}\right\rangle_{\text {odd, } \mathbf{h}}=\left(k_{1} k_{2} k_{3}\right)^{s-1} \frac{1}{E^{3}}\left[\left\{\left(\vec{k}_{1} \cdot \vec{z}_{3}\right)\left(\epsilon^{k_{3} z_{1} z_{2}} k_{1}-\epsilon^{k_{1} z_{1} z_{2}} k_{3}\right)+\left(\vec{k}_{3} \cdot \vec{z}_{2}\right)\left(\epsilon^{k_{1} z_{1} z_{3}} k_{2}-\epsilon^{k_{2} z_{1} z_{3}} k_{1}\right)\right.\right. \notag\\
&\left.\left.-\left(\vec{z}_{2} \cdot \vec{z}_{3}\right) \epsilon^{k_{1} k_{2} z_{1}} E+\frac{k_{1}}{2} \epsilon^{z_{1} z_{2} z_{3}} E\left(E-2 k_{1}\right)\right\}+\text { cyclic perm }\right] \notag\\
&\times\left[\frac{1}{E^{3}}\left\{2\left(\vec{z}_{1} \cdot \vec{k}_{2}\right)\left(\vec{z}_{2} \cdot \vec{k}_{3}\right)\left(\vec{z}_{3} \cdot \vec{k}_{1}\right)+E\left\{k_{3}\left(\vec{z}_{1} \cdot \vec{z}_{2}\right)\left(\vec{z}_{3} \cdot \vec{k}_{1}\right)+\text { cyclic }\right\}\right\}\right]^{s-1}
\end{align}
which in the flat-space limit $E \to 0$ becomes
\begin{align}
&\lim_{E\to 0}\left\langle J_{s} J_{s} J_{s}\right\rangle_{\text {e }, \mathbf{h}}\sim (k_1k_2k_3)^{s-1}(z_1.k_2)^s(z_2.k_3)^s(z_3.k_1)^s\notag\\
&\lim_{E\to 0}\left\langle J_{s} J_{s} J_{s}\right\rangle_{\text {o}, \mathbf{h}} \sim (k_1k_2k_3)^{s-1}[\epsilon(z_1z_2k_1)k_2-\epsilon(z_1z_2k_2)k_1](z_3.k_1)^s(z_1.k_2)^{s}(z_2.k_3)^{s}
\end{align} where in the last line we have used \eqref{3d2d} to convert the amplitude to four dimensional language.
As usual, the minimal part of $\mathcal{A}^{sss}_{even}$ and $\mathcal{A}^{sss}_{odd}$ appear in the flat-space limit.
One does similar computation for the non-homogeneous piece. In general non-homogeneous pieces are very complicated. However, we can look into a few examples such as $\langle TTT \rangle_{nh}$. We summarize the findings in a few tables.

\begin{table}[h!]
\begin{tabular}{||c | c | c ||}
\hline
\multicolumn{3}{|c|}{Inside the triangle $\langle J_{s_1}J_{s_2}J_{s_3}\rangle$(Exactly conserved current)}\\
\hline\hline
\begin{tabular}{@{}c@{}}\bf{4D flat-space} \\ \bf{Amplitudes}\end{tabular} & \bf{3D CFT correlator} &\begin{tabular}{@{}c@{}}\bf{Expected CFT} \\ \bf{Pole structure}\end{tabular} \\ [0.5ex]
\hline
Non-minimal (Even) & Homogenous & $E^{-s_1-s_2-s_3}$ \\
\hline
Minimal (Even) & Non-Homogenous & $E^{-s_1-s_2-s_3+2}$ \\
\hline
Non-minimal (Odd)& Homogenous & $E^{-s_1-s_2-s_3}$ \\
\hline
Minimal (Odd) for $s_1\ne s_2\ne s_3$ & \begin{tabular}{@{}c@{}} No corresponding \\ CFT correlator\end{tabular} & $\cross$ \\ [1ex]
\hline
\end{tabular}
\caption{Flat-space limit of CFT correlator.}
\label{T1}
\end{table}

\subsection{Outside triangle inequality}
Let us now consider a few examples of correlators outside the triangle. One very important aspect which is different for correlators outside the triangle as compared to inside is that all the correlator outside the triangle is non-homogeneous \cite{Jain:2021whr}.
\subsubsection*{$\langle J_s O O \rangle$}
In momentum space, we have
\begin{align}
\left\langle J_{s} O_{\Delta} O_{\Delta}\right\rangle_{\text {e }}=c_{1} k_{1}^{2 s-1} I_{\frac{1}{2}+s,\left\{\frac{1}{2}-s_{1} \Delta-\frac{3}{2}, \Delta-\frac{3}{2}\right\}}\left(k_{2} \cdot z_{1}\right)^{s}
\end{align}
which in the flat-space limit $E \to 0$ becomes
\begin{align}
\lim_{E\to 0}\left\langle J_{s} O_{\Delta} O_{\Delta}\right\rangle_{\text {even }}\sim \frac{k^{s-1}_1 (k_2k_3)^{\Delta-2}}{E^s}(z_1.k_2)^s
\end{align}
where again we see $\mathcal{A}^{100}_{e}$ appear in the flat-space limit.
\subsection*{$\langle J_3J_1O\rangle$}
The $\langle J_3J_1O\rangle$ correlator in momentum space is given by
\begin{align}
& \langle J_3J_1O\rangle = \left(-\frac{E^2+2k_1(E+k_1)}{E^4}\right)(z_1.k_2)^3 z_2.k_1\notag\\&+\left(\frac{E^3-2k^2_1(E+2k_1)-2(3 E^2 + 3 E k_1 + 2k^2_1)k_2}{2E^3}\right)(z_1.k_2)^2 z_1.z_2
\end{align}
Now, we take the flat-space limit $E \to 0$, where we see that
\begin{align}
\lim_{E\to 0}\langle J_3J_1O\rangle = \frac{2k^2_1}{E^4}(z_1\cdot k_2)^3z_2\cdot k_1
\end{align}which matches with flat-space amplitude.
The parity-odd term can be obtained from free fermion theory and teh amplitude can be identified in a similar way.

It is interesting to note that correlators involving exactly conserved currents outside the triangle do not have any parity-odd contribution \cite{Giombi:2011rz}. To get the parity-odd non-minimal coupling, we need to consider theories with weakly broken Higher-Spin symmetry \cite{Giombi:2016zwa} which we consider in the next section.

The general correspondence is listed below.
\begin{table}[h!]
\begin{tabular}{||c | c | c ||}
\hline
\multicolumn{3}{|c|}{Outside the triangle $\langle J_{s_1}J_{s_2}J_{s_3}\rangle$(Exactly conserved current)}\\
\hline\hline
\begin{tabular}{@{}c@{}}\bf{4D Flat-space} \\ \bf{Amplitudes}\end{tabular} & \bf{3D CFT correlator} & \begin{tabular}{@{}c@{}}\bf{Expected CFT} \\ \bf{Pole structure}\end{tabular}\\ [0.5ex]
\hline
Non-minimal (Even) & Non-Homogenous $(b-f)$ & $E^{-s_1-s_2-s_3}$ \\
\hline
Minimal (Even) & Non-Homogenous $(b+f)$ & $E^{-s_1-s_2-s_3+2}$ \\
\hline
Non-minimal (Odd)& \begin{tabular}{@{}c@{}}{\bf {No corresponding}} \\ {\bf {CFT correlator}}\end{tabular}& $\cross$ \\
\hline
Minimal (Odd) & \begin{tabular}{@{}c@{}}{\bf {No corresponding}} \\ {\bf {CFT correlator}}\end{tabular} & $\cross$ \\ [1ex]
\hline
\end{tabular}
\caption{Correspondence outside the triangle inequality. Here $b-f$ implies subtraction of free bosonic correlator and free fermionic correlator. Let us also note that there is no parity-odd CFT correlator for exactly conserved current outside the triangle.}
\label{T1}
\end{table}

\begin{table}[h!]
\begin{tabular}{||c | c | c ||}
\hline
\multicolumn{3}{|c|}{Outside the triangle $\langle OJ_{s_2}J_{s_3}\rangle$(Exactly conserved current)}\\
\hline\hline
\begin{tabular}{@{}c@{}}\bf{4D flat-space} \\ \bf{Amplitudes}\end{tabular} & \bf{3D CFT correlator} & \begin{tabular}{@{}c@{}}\bf{Expected CFT} \\ \bf{Pole structure}\end{tabular} \\ [0.5ex]
\hline
\begin{tabular}{@{}c@{}}Even\end{tabular}& Non-Homogeneous free bosonic &$E^{-s_2-s_3}$ \\
\hline
\begin{tabular}{@{}c@{}}Odd \end{tabular}& Non-Homogeneous free fermionic & $E^{-s_2-s_3}$ \\ [1ex]
\hline
\end{tabular}
\caption{parity-even and odd part of the amplitude can be obtained from the flat-space limit of the free bosonic and free fermionic CFT correlator respectively.}
\label{T11}
\end{table}

\begin{table}[h!]
\begin{tabular}{||c | c | c ||}
\hline
\multicolumn{3}{|c|}{Outside the triangle $\langle J_{s}OO\rangle$(Exactly conserved current)}\\
\hline\hline
\begin{tabular}{@{}c@{}}\bf{4D Flat-space} \\ \bf{Amplitudes}\end{tabular} & \bf{3D CFT correlator} & \begin{tabular}{@{}c@{}}\bf{Expected CFT} \\ \bf{Pole structure}\end{tabular}\\ [0.5ex]
\hline
\begin{tabular}{@{}c@{}}Even\end{tabular}& Homogenous & $E^{-s}$\\ [1ex]
\hline
\end{tabular}
\caption{For this case there is only one structure and can be obtained by considering free bosonic or free fermionic theory alone.}
\label{T12}
\end{table}

\subsection*{Weakly broken Higher-spin theory}
As discussed above, there is no parity-odd contribution to the correlator outside the triangle for exactly conserved currents. However, it is known that for weakly broken currents it is possible to have a parity-odd contribution for such cases \cite{Maldacena:2012sf,Giombi:2016zwa}. Recently, in \cite{Jain:2021whr}, the parity-odd part was computed in momentum space. Let us consider $\langle J_1 J_1 J_4\rangle$ for simplicity. This correlator was computed explicitly in \cite{Jain:2021whr}.
The parity-odd piece is given by a non-homogeneous piece
\begin{align}
&\langle J_1 J_1 J_4\rangle_{o,nh_1} = ([O_{\epsilon}]_1+[O_{\epsilon}]_2)\langle J_1 J_1 J_4\rangle_{b-f}.
\end{align}
It is easy to show that in the flat-space limit this gives rise to non-minimal parity-odd amplitude
\begin{align}
&\lim_{E\to 0}([O_{\epsilon}]_1+[O_{\epsilon}]_2)\langle J_1 J_1 J_4\rangle_{b- f} \sim [\mathcal{M}^{114}_{o}]_{nm}.
\end{align}

\begin{table}[h!]
\begin{tabular}{||c | c | c ||}
\hline
\multicolumn{3}{|c|}{Outside the triangle (Slightly broken higher spin symmetry)}\\
\hline\hline
\begin{tabular}{@{}c@{}}\bf{4D flat-space} \\ \bf{Amplitudes}\end{tabular} & \bf{3D CFT correlator} & \begin{tabular}{@{}c@{}}\bf{Expected CFT} \\ \bf{Pole structure}\end{tabular} \\ [0.5ex]
\hline
Non-minimal (Even) & Non-Homogenous $(b-f)$ & $E^{-s_1-s_2-s_3}$\\
\hline
Minimal (Even) &Non-Homogenous $(b+f) $ & $E^{-s_1-s_2-s_3+2}$ \\
\hline
Non-minimal (Odd)& $s_1\neq 0$, $\langle J_{s_1}J_{s_2} J_{s_3}\rangle_{odd, nh_1}$ & $E^{-s_1-s_2-s_3}$\\
\hline
Minimal (Odd) & No corresponding correlator & $\cross$ \\ [1ex]
\hline
\end{tabular}
\caption{Flat-space limits for outside the triangle correlators.}
\label{T3}
\end{table}

\section{Flat space limit in spinor helicity variable}\label{fltsp1}
The most general form of the CFT correlator in spinor helicity is given by
\begin{align}
\left\langle J_{s_{1}}^{h_{1}} J_{s_{2}}^{h_{2}} J_{s_{3}}^{h_{3}}\right\rangle=f_{h_{1}, h_{2}, h_{3}}\left(k_{1}, k_{2}, k_{3}\right)\langle 12\rangle^{h_{3}s_3-h_{1}s_1-h_{2}s_2}\langle 23\rangle^{h_{1}s_1-h_{2}s_2-h_{3}s_3}\langle 31\rangle^{h_{2}s_2-h_{3}s_3-h_{1}s_1}\label{old}
\end{align}
However, for clarity let us consider two different class of CFT structures based on their net helicities $H$ and rewrite the most general CFT correlator in spinor-helicity as
\begin{align}
\left\langle J_{s_{1}}^{h_{1}} J_{s_{2}}^{h_{2}} J_{s_{3}}^{h_{3}}\right\rangle= \begin{cases}
f_{h_1, h_2, h_3}\langle 12\rangle^{h_{3}s_3-h_{1}s_1-h_{2}s_2}\langle 31\rangle^{h_{2}s_2-h_{3}s_3-h_{1}s_1}\langle 23\rangle^{h_{1}s_1-h_{2}s_2-h_{3}s_3} & ~~H<0 \\ f_{h_1, h_2, h_3} \langle \bar{1}\bar{2}\rangle^{h_{1}s_1+h_{2}s_2-h_{3}s_3}\langle\bar{3}\bar{1}\rangle^{h_{3}s_3+h_{1}s_1-h_{2}s_2}\langle\bar{2}\bar{3}\rangle^{h_{2}s_2+h_{3}s_3-h_{1}s_1} & ~~H>0
\end{cases}\label{new}
\end{align}
where $H = h_{1}s_1+h_{2}s_2+h_{3}s_3.$
We also have
\begin{align}
\langle ij\rangle\langle\bar{i}\bar{j} \rangle = -E(k_i+k_j-k_k)
\end{align} in three dimensional CFT whic implies
the barred spinor brackets and the unbarred spinor brackets are related, therefore, (\ref{old}) and (\ref{new}) are equivalent\footnote{Except when $h_{1}s_1+h_{2}s_2+h_{3}s_3=0.$ This case discussed below separately.}. In the limit of $E\rightarrow 0,$, we see that
\begin{align}
\langle ij\rangle\langle\bar{i}\bar{j} \rangle = 0
\end{align}
If the momenta are real, then $\langle ij\rangle^{*} = \langle\bar{i}\bar{j} \rangle$, then both barred and unbarred brackets are zero. If the momenta are complex, then the barred brackets and the unbarred brackets are independent and in the flat-space limit only one can be non-zero i.e $\langle\bar{i}\bar{j} \rangle = 0$ or $\langle ij\rangle = 0$. Also in the flat space limit, barred and un barred brackets are independent of each other. One can see that the flat-space limit of (\ref{new}) reproduces (\ref{nonpamp}) where $\langle\bar{i}\bar{j} \rangle $ must be identified with $[ij]$, therefore,
\begin{align}
\lim_{E\to 0}\left\langle J_{s_{1}}^{h_{1}} J_{s_{2}}^{h_{2}} J_{s_{3}}^{h_{3}}\right\rangle = \mathcal{A}^{s_{1}, s_{2}, s_{3}}_{h_1, h_2, h_3}
\end{align}
Let us see the above happen with the example of $\langle JJO\rangle$ which in spinor-helicity variables is given by
\begin{align}
&\langle J_{-}J_{-}O\rangle_{even} = \frac{\langle 12\rangle^2}{E^2} \quad \langle J_{-}J_{-}O\rangle_{odd} = i\frac{\langle 12\rangle^2}{E^2}\quad H < 0\\
&\langle J_{+}J_{+}O\rangle_{even} = \frac{\langle \bar{1}\bar{2}\rangle^2}{E^2} \quad \langle J_{+}J_{+}O\rangle_{odd} = -i\frac{\langle \bar{1}\bar{2}\rangle^2}{E^2}\quad H > 0\\
&\langle J_{+}J_{-}O\rangle_{even} = \langle J_{+}J_{-}O\rangle_{odd} = 0\quad ~ \rm{for}~ H = 0
\end{align}
In the flat space limit we obtain
\begin{align}
&\lim_{E\to 0}\langle J_{-}J_{-}O\rangle =\frac{\langle 12\rangle^2}{E^2}= \frac{\mathcal{A}^{110}_{--}}{E^2} \quad H < 0\\
&\lim_{E\to 0}\langle J_{+}J_{+}O\rangle = \frac{[ 12 ]^2}{E^2}= \frac{\mathcal{A}^{110}_{++}}{E^2} \quad H > 0\\
&\lim_{E\to 0}\langle J_{+}J_{-}O\rangle= 0 \quad H = 0
\end{align}
which precisely matches all the helicity components of $\mathcal{A}^{110}$.

\section{More on new parity odd CFT correlator}\label{npodapd}
In this Appendix we consider another example of new parity odd CFT correlator proposed in section \ref{nptod}. We also discuss the same in spinor helicity variables.
\subsection{$\langle J_1J_1 J_4\rangle$}
In the momentum space the CFT correlator \footnote{Correlators which are out-side the triangle, has only non-homogeneous piece. Here we talk about nh component obtained by adding free boson and free fermion $\langle J_4J_1J_1\rangle$ correlator.} is given by \cite{Jain:2021whr}
\begin{align}
& \langle J_1J_1J_4\rangle_{nh,b+f} =\notag\\& \left(-\frac{5 E^3+k_3(5E^2+2k_3(2E+k_3))}{512 E^4}\right)[(z_3.k_2)^4 z_2.z_1 + (z_3\cdot k_2)^3 z_3.z_2 z_1\cdot k_3 - (z_3\cdot k_2)^3 z_3\cdot z_1 z_2\cdot k_3]
\end{align}
Now, we take the flat-space limit $E \to 0$, where we see that
\begin{align}
\lim_{E\to 0}\langle J_1J_1J_4\rangle_{b+f} = \frac{k^3_3}{256 E^4}[(z_3.k_2)^4 z_2.z_1 + (z_3\cdot k_2)^3 z_3.z_2 z_1\cdot k_3 - (z_3\cdot k_2)^3 z_1\cdot z_3 z_2\cdot k_3]
\end{align}
which is precisely the parity-even minimal amplitude obtained in (\ref{114ampa}). Now, we look at the difference
\begin{align}
\langle J_1J_1J_4\rangle_{b-f} = \left(-\frac{5 E^3+k_3(15E^2+4k_3(6E+5k_3))}{2560 E^6}\right)(z_3.k_2)^4 z_2.k_1 z_1.k_3 + \mathcal{O}(\frac{1}{E^5})
\end{align}
which in the flat-space limit $E\to 0$ gives
\begin{align}
\lim_{E\to 0}\langle J_1J_1J_4\rangle_{b-f} = -\frac{k^3_3}{128 E^6}(z_3.k_2)^4 z_2.k_1 z_1.k_3
\end{align}
which is precisely the parity-even non-minimal amplitude obtained in (\ref{114ampa}). To obtain the parity-odd vertex, we consider the epsilon transform of the above. Since, this correlator is outside the triangle, there will be no parity-odd contribution for exactly conserved currents, therefore, we consider weakly broken current correlation to generate parity-odd terms. The parity-odd contributions in momentum space were computed explicitly in \cite{Jain:2021whr}. Using parity odd result in \cite{Jain:2021whr} it is easy to show that
\begin{align}\label{od1a}
&\lim_{E\to 0}([O_{\epsilon}]_1+[O_{\epsilon}]_2)\langle J_{1}J_{1}J_{4}\rangle_{b-f} \sim \mathcal{M}^{114}_{nm, o}
\end{align}
Using \eqref{goutnh1} we can define parity odd CFT correlator which in the flat space limit gives parity odd minimal amplitude. More precisely,
\begin{align}\label{od1ab}
&\lim_{E\to 0}([O_{\epsilon}]_1-[O_{\epsilon}]_2)\langle J_{1}J_{1}J_{4}\rangle_{b+f} \sim \mathcal{M}^{114}_{m, o}
\end{align}
generates the desired parity-odd minimal cubic vertex in (\ref{114ampa}). The anti-symmetry is due to the Chan-Paton factors discussed earlier near (\ref{sss3}).

We now consider spinor helicity variables to describe the new parity odd CFT correlators. This will give us some more understanding of the same correlator.

\subsection{In spinor helicity variables}
Let us consider $\langle JJT\rangle$ in the spinor-helicity variables
\begin{align}\label{112spans}
&\langle J_{-}J_{-}T_{-}\rangle = \left(g_e+ i g_o\right)\frac{\langle 23\rangle^2\langle 31\rangle^2 k_3}{E^4}\nonumber\\
&\langle J_{-}J_{-}T_{-}\rangle =\langle J_{-}J_{-}T_{+}\rangle_{nh} = 0\nonumber\\
&\langle J_{+}J_{-}T_{-}\rangle = c_J\frac{\langle 23\rangle^4 (E-2k_3)^2(E+k_3)}{\langle 12\rangle^2 k^2_3E^2}
\end{align}
and its conjugate.\footnote{Here the conjugate is a means to generate the barred spinors bracket and doesn't actually mean the complex conjugate of the CFT correlators. For example, the conjuagte of $\langle T^{-}T^{-}T^{-}\rangle$ is $\langle T^{+}T^{+}T^{+}\rangle \sim \langle \bar{1}\bar{2}\rangle^{2}\langle \bar{2}\bar{3}\rangle^{2}\langle \bar{3}\bar{1}\rangle^{2}$}
Since, in the flat space limit, we have
\begin{align}
\langle ij\rangle \langle \bar{i}\bar{j}\rangle = E(k_i+k_j-k_k) \to 0
\end{align}
Therefore, either $\langle ij\rangle = 0$ or $\langle \bar{i}\bar{j}\rangle = 0$ in the flat-sapce limit. If we work with $\langle \bar{i}\bar{j}\rangle = 0$\footnote{ $\langle \bar{i}\bar{j}\rangle = [ij]$}, then
\begin{align}\label{211spiamp}
&\lim_{E\to 0}\langle J_{-}J_{-}T_{-}\rangle = \frac{\langle 23\rangle^2\langle 31\rangle^2 k_3}{E^4}\equiv \frac{[\mathcal{M}_{nm}]^{112}_{---}k_3}{E^4}\nonumber\\
&\lim_{E \to 0}\langle J_{+}J_{-}T_{-}\rangle = c_J\frac{\langle 23\rangle^4 k_3}{\langle 12\rangle^2 E^2} +c_J O(\frac{1}{E^2}) \equiv \frac{[\mathcal{M}_{m}]^{112}_{+--}k_3}{E^2}
\end{align}
where we have identified with the flat-space photon-photon-graviton vertices computed in (\ref{ppgcubv}). If we work with $\langle \bar{i}\bar{j}\rangle = 0$ then
\begin{align}\label{112msmt}
&\lim_{E\to 0}\langle J_{+}J_{+}T_{+}\rangle = \frac{\langle\bar{ 2}\bar{3}\rangle^2\langle \bar{3}\bar{1}\rangle^2 k_3}{E^4}\equiv \frac{[\mathcal{M}_{nm}]^{112}_{+++}k_3}{E^4}\nonumber\\
&\lim_{E \to 0}\langle J_{-}J_{+}T_{+}\rangle = c_J\frac{\langle \bar{2}\bar{3}\rangle^4 k_3}{\langle \bar{1}\bar{2}\rangle^2 E^2} +c_J O(\frac{1}{E^2})\equiv \frac{[\mathcal{M}_{m}]^{112}_{-++}k_3}{E^4}
\end{align}
where again we have identified the flat-space photon-photon-graviton vertices in (\ref{ppgcubv}). One can also convert \eqref{TJJnho} in spinor helicity variables and show that it is consistent with (\ref{ppgcubv}).

It is interesting to note that, if we ignore net helicity zero amplitude, then there is no mismatch between spinor helicity \eqref{ppgamp} and cubic vertex \eqref{pp4cubv} answers. Similarly, for the CFT correlator in spinor helicity variables, there is no net helicity zero CFT correlation which is consistent with locality. See appendix \ref{zeroamp} for more discussion.

\subsection*{$\langle TTT\rangle$ in spinor helicity variables}
Let us now look at $\langle TTT\rangle$ in spinor-helicity\footnote{The contact term in \eqref{contnhod} takes the form
	\begin{align}\label{tttoddnhab}
	&\left\langle T^{-} T^{-} T^{-}\right\rangle=c_{T}^{\prime} \frac{E^{3}-E b_{123}-c_{123}}{c_{123}^{2}}\langle 12\rangle^{2}\langle 23\rangle^{2}\langle 31\rangle^{2} \notag\\
	&\left\langle T^{-} T^{-} T^{+}\right\rangle=c_{T}^{\prime} \frac{[\left(E-2 k_{3}\right)^{3}-\left(E-2 k_{3}\right)\left(b_{123}-2 k_{3} a_{12}\right)+c_{123}](E-2k_2)^2(E-2k_1)^2}{c_{123}^{2}}\frac{\langle 12\rangle^{6}}{\langle 31\rangle^{2}\langle 23\rangle^{2}}
	\end{align}
} variables
\begin{align}\label{tttsph1}
&\left\langle T^{-} T^{-} T^{-}\right\rangle=\left(\left(c_{1}+i c_1'\right) \frac{c_{123}}{E^{6}}+c_{T} \frac{E^{3}-E b_{123}-c_{123}}{c_{123}^{2}}\right)\langle 12\rangle^{2}\langle 23\rangle^{2}\langle 31\rangle^{2} \notag\\
&\left\langle T^{-} T^{-} T^{+}\right\rangle=c_{T} \frac{\left(E-2 k_{3}\right)^{2}\left(E^{3}-E b_{123}-c_{123})\right(E-2k_2)^2(E-2k_1)^2}{E^{2} c_{123}^{2}}\frac{\langle 12\rangle^{6}}{\langle 31\rangle^2\langle 23\rangle^2}
\end{align}
its conjugate and cyclic permutation. We have not displayed parity-even or parity-odd correlators separately. The parity odd homogeneous part is proportional to $ c_1'.$ The term proportional to $c_T$ is parity even non-homogeneous and does not have parity odd analogue for conserved current. In the flat space limit we have
\begin{align} \label{tttns}
&\lim_{E\to 0}\left\langle T^{-} T^{-} T^{-}\right\rangle = \left(c_1+i c_1'\right)\frac{ \langle 12\rangle^{2}\langle 23\rangle^{2}\langle 31\rangle^{2}}{E^6}+\rm{subleading~terms} \nonumber\\
&\lim_{E\to 0}\left\langle T^{-} T^{-} T^{+}\right\rangle = c_T\frac{k_1k_2k_3}{E^2}\frac{\langle 12\rangle^{6}}{\langle 31\rangle^{2}\langle 23\rangle^{2}} +{\mathcal{O}\left(\frac{1}{E}\right)}
\end{align}
which matches with both cubic vertex and spinor helicity amplitude \eqref{222cv}, \eqref{gggamp}.
To produce correct flat space parity odd spinor helicity minimal amplitude \footnote{Interestingly \eqref{tttoddnhab} has the missing the terms in spinor helicity variables
	which can be identified with $\mathcal{A}_{--+}^{222}$ that appears in \eqref{gggamp}. However, this is a contact term and can not be thought as reproducing flat space amplitude.} we look at (\ref{anoddttt}) in the spinor-helicity variables
\begin{align}
&\left\langle T^{-} T^{-} T^{-}\right\rangle'_{\text {o}}=i\left( \frac{E^{3}-E b_{123}-c_{123}}{c_{123}^{2}}\right)\langle 12\rangle^{2}\langle 23\rangle^{2}\langle 31\rangle^{2} \notag\\
&\left\langle T^{-} T^{-} T^{+}\right\rangle'_{\text {o}}=i \frac{\left(E-2 k_{3}\right)^{2}\left(E^{3}-E b_{123}-c_{123})\right(E-2k_2)^2(E-2k_1)^2}{E^{2} c_{123}^{2}}\frac{\langle 12\rangle^{6}}{\langle 31\rangle^2\langle 23\rangle^2}
\end{align}
which in the flat space limit gives
\begin{align}
\lim_{E \to 0}\langle T^{-}T^{-}T^{+}\rangle'_{o} \sim \frac{1}{E^2}\frac{\langle 12\rangle^{6}}{\langle 31\rangle^{2}\langle 23\rangle^{2}} +{\mathcal{O}\left(\frac{1}{E}\right)}
\end{align}
which matches precisely with missing parity odd term $\mathcal{A}_{--+}^{222}$ in
(\ref{gggamp}). This analysis of $\langle TTT\rangle$ can be generalized for any arbitrary equal spin correlator $\langle J_s J_s J_s\rangle.$ So we conclude that, for equal spin case, we are able to construct CFT correlator which in the flat space limit reproduces correct amplitude in spinor helicity variables.

\subsection*{$\langle J_1J_1 J_4\rangle$ in spinor helicity variables}

Consider the correlator $\langle J_1J_1J_4\rangle$ in spinor-helicity, in two separate linear combinations
\begin{align}
&\langle J_{-}J_{-}J_{4-}\rangle_{b-f} = \langle 32\rangle^4\langle 31\rangle^4 \langle\bar{2}\bar{1}\rangle^2\frac{3E^5+5E^4 k_3+8 E^3 k^2_3+ 12 E^2 k^3_3+ 16 E k^4_3 + 16 k^5_3}{E^8 k^4_3}\nonumber\\
&\langle J_{-}J_{-}J_{4+}\rangle_{b-f} = \langle 21\rangle^6\langle\bar{3}\bar{2}\rangle^4\langle\bar{1}\bar{3}\rangle^4\frac{3E+k_3}{E^8 k^4_3}\nonumber\\
&\langle J_{+}J_{-}J_{4-}\rangle_{b-f} = 0
\end{align}
Parity odd results in \eqref{od1a} in spinor helicity variable is identical to above answer in spinor helicity upto some factor of $i.$
and the other combination
\begin{align}
&\langle J_{-}J_{-}J_{4-}\rangle_{b+f} = \langle J_{-}J_{-}J_{4+}\rangle_{b+f} = 0 \nonumber\\
&\langle J_{+}J_{-}J_{4-}\rangle_{b+f} = \langle 23\rangle^6\langle\bar{1}\bar{2}\rangle^4\langle\bar{2}\bar{3}\rangle^4\frac{5E^3+5E^2k_3+4E k^2_3+2k^3_3}{E^8 k^4_3}
\end{align}
and its conjugate. Parity odd results in \eqref{od1ab} in spinor helicity variable is identical to above answer in spinor helicity upto some factor of $i.$ We look at the flat-space limits of the above
\begin{align}
&\lim_{E\to 0}\langle J_{-}J_{-}J_{4-}\rangle_{b-f} = \frac{\langle 32\rangle^4\langle 31\rangle^4}{\langle 12\rangle^2} \frac{16 k^3_3}{E^6}+ O(\frac{1}{E^5})\label{eqab}\nonumber\\
&\lim_{E\to 0}\langle J_{-}J_{-}J_{4+}\rangle_{b-f} \approx 0\nonumber\\
&\lim_{E\to 0}\langle J_{+}J_{-}J_{4-}\rangle_{b-f} = 0
\end{align}
where $\approx$ means that there are no singularities in $E$. Now let us look at the other combination
\begin{align} \label{eqab1}
&\lim_{E\to 0}\langle J_{-}J_{-}J_{4-}\rangle_{b+f} = \lim_{E\to 0}\langle J_{-}J_{-}J_{4+}\rangle_{b+f} = 0 \notag\\
&\lim_{E\to 0}\langle J_{+}J_{-}J_{4-}\rangle_{b+f} = \frac{\langle 23\rangle^6\langle 31\rangle^2}{\langle 12\rangle^4}\frac{k^3_3}{E^4}+O(E^0)
\end{align}
Notice that the flat-space limits correspond to the photon-photon-$spin_4$ cubic vertices (\ref{pp4cubv}).
We observe that in both the combination, in the leading order there is no contribution in $--+$ or $++-$ helicity\footnote{However, if we include non-singular contributions as well, namely in(\ref{eqab1}),
	\begin{align}
	&\lim_{E\to 0}\langle J_{-}J_{-}J_{4+}\rangle_{b-f} = \frac{\langle 21\rangle^6}{\langle 23\rangle^4\langle 31\rangle^4}\frac{k^4_1 k^4_2}{k^3_3}
	\end{align}
	We can see that the above can be identified with $\mathcal{A}^{114}_{--+}$ and therefore, the full correlator reproduces the angle-brackets of (\ref{pp4amp}) while the conjugate correlator reproduces the square-brackets.}.
We have seen that covariant vertex answer in \eqref{cubv}, the helicity component ${\mathcal M}_{--+}={\mathcal M}_{++-}=0$ whereas spinor helicity amplitude ${\mathcal A}_{--+}\neq 0,{\mathcal A}_{++-}\neq 0.$ In \eqref{eqab} we ignored subleading terms as they are not singular in the limit $E\rightarrow 0.$ Neglecting subleading terms the flat space limit, we conclude that the CFT correlator $\langle J_1J_1 J_4\rangle$ reproduces correct flat space covariant vertex but not the full spinor helicity amplitude in \eqref{cubv}.
This implies that in general flat space limit of the CFT correlator, in general, reproduces the correct cubic vertex but not the full spinor helicity amplitude.
\section{Constructing missing CFT correlator for $\langle JJ J_4\rangle$}\label{411ab}
If we ignore subleading terms, it is shown in \eqref{eqab}, \eqref{eqab1} that $\langle JJ J_4\rangle$
does not produce $--+$ or $++-$ helicity amplitude in the flat space limit.

It turns out that one can construct more homogeneous solutions such that
\begin{align}
K^{\kappa} \frac{ \langle J^{-}J^{-} J_4^{+}\rangle }{k_3} = 0, \quad K^{\kappa} \frac{ \langle J^{+}J^{+} J_4^{-}\rangle }{k_3} = 0.
\end{align}

One can explicitly solve the above equation to obtain
\begin{align}
&\langle J_{1-}J_{1-}J_{4+}\rangle = 
E^2\frac{\langle 12\rangle^6}{\langle 23\rangle^4\langle 31\rangle^4}
\end{align} and its complex conjugate.
Converting the into momentum space gives
\begin{align}
&\langle J_{1}(z_1, k_1)J_{1}(z_2, k_2) J_4(z_3, k_3)\rangle = A(z_3.k_2)^4 z_2.k_1 z_1.k_3 + B z_1.z_2 (z_3.k_2)^4 + C(z_3.z_2)(z_3.k_2)^3(z_1.k_3)\notag\\&+C(k_2 \leftrightarrow k_1) z_3.z_1 (z_3.k_2)^3(z_2.k_1)
\end{align}
where
\begin{align}
&A = \frac{8 k^6_3}{E(E-2k_1)^4(E-2k_2)^4} ,~~~
B = -\frac{4(E-2k_3)k^7_3}{E(E-2k_1)^4(E-2k_2)^4},~~
&C = \frac{4k_1k^6_3 (E-2k_3)}{E(E-2k_1)^4(E-2k_2)^4}.
\end{align}
It is easy to see that this CFT correlator has a bad pole, that is it has poles at other momentum configuration other than $E \rightarrow 0.$

One can check that the flat space limit cannot be re-written as $4D$ Lorentz invariant structure henec even though it reproduces correct flat space spinor helicity amplitude, it does not lead to any new covariant vertex.
\section{Various-Identities}\label{idnty}
In this section, we derive
\begin{align}
&Y_2Y_3 [O_{\epsilon}]_2G = -GV_1 \label{SID4}\\
&Y_3[O_{\epsilon}]_2Y_2 = -V_1\label{SID5}\\
&Y_2Y_3 [O_{\epsilon}]_1G = GV_1 \label{SID6}
\end{align}
Consider first
\begin{align}
Y_2Y_3[O_{\epsilon}]_2G &= (z_2.k_3)(z_3.k_1)\left[(z_1.k_2)\frac{\epsilon(z_2k_2z_3)}{k_2}+(z_3.z_1)\frac{\epsilon(z_2k_2k_3)}{k_2}+(z_3.k_1)\frac{\epsilon(z_2k_2z_1)}{k_2}\right]\\
&=(z_3.k_1)\left[(z_1.k_2)(z_2.z_3)\frac{\epsilon(z_2k_2k_3)}{k_2}+(z_3.z_1)(z_2.k_3)\frac{\epsilon(z_2k_2k_3)}{k_2}+(z_3.k_1)(z_1.z_2)\frac{\epsilon(z_2k_2k_3)}{k_2}\right]
\end{align}
where in the second equality we take $z_2.k_3$ inside the bracket and make use of the Schouten identities
\begin{align}
&\epsilon(z_2k_2z_3)Y_2=\epsilon(z_2k_2z_3)(z_2.k_3) = \epsilon(z_2k_2k_3)(z_2.z_3) = \epsilon(z_2k_2k_3) Z_1\\
&\epsilon(z_2k_2z_1)Y_2=\epsilon(z_2k_2z_1)(z_2.k_3) = \epsilon(z_2k_2k_3)(z_1.z_2) = \epsilon(z_2k_2k_3)Z_3
\end{align}
to obtain
\begin{align}
Y_2Y_3[O_{\epsilon}]_2G &=(z_3.k_1)\frac{\epsilon(z_2k_2k_3)}{k_2}G\label{S2}
\end{align}
Now, consider the schouten identity
\begin{align}
\epsilon(z_2k_2k_3) (z_3.k_2) = -\epsilon(z_2 z_3 k_2) k_2.k_3+\epsilon(z_2z_3k_3)k^2_2
\end{align}
since, $p_I.p_J = k_Ik_J-k_I.k_J = 0$, the above schouten identity becomes
\begin{align}
\epsilon(z_2k_2k_3) (z_3.k_2) = -\epsilon(z_2 z_3 k_2) k_2k_3+\epsilon(z_2z_3k_3)k^2_2 = -k_2\epsilon(z_2z_3p_2p_3) = k_2V_1\label{SID61}
\end{align}
which we now use in (\ref{S2}) to get
\begin{align}
Y_2Y_3[O_{\epsilon}]_2G &=-GV_1
\end{align}
Similarly one can show
\begin{align}
Y_2Y_3[O_{\epsilon}]_3G &=-GV_1
\end{align}
Similarly, we derive (\ref{SID5}). Consider the epsilon transform of $Y_2$ as follows
\begin{align}
Y_3[O_{\epsilon}]_2Y_2 =(z_3.k_1) \frac{\epsilon(z_2k_2k_3)}{k_2}
\end{align}
Now, by using (\ref{SID61}) in the above, we immediately see
\begin{align}
Y_3[O_{\epsilon}]_2Y_2 = -V_1
\end{align}
Similarly, one can show
\begin{align}
Y_2[O_{\epsilon}]_3Y_3 = -V_1
\end{align}
To derive (\ref{SID6}), consider
\begin{align}
Y_2Y_3[O_{\epsilon}]_2G &= (z_2.k_3)(z_3.k_1)\left[(z_2.z_3)\frac{\epsilon(z_1k_1k_2)}{k_1}+(z_2.k_3)\frac{\epsilon(z_1k_1z_3)}{k_1}+(z_3.k_1)\frac{\epsilon(z_1k_1z_2)}{k_1}\right]
\end{align}
Now, we take $z_3.k_1$ inside the bracket and use the following Schouten identities
\begin{align}
&Y_3 \epsilon(z_1k_1z_2) = (z_3.k_1) \epsilon(z_1k_1z_2) = -k^2_1\epsilon(z_1z_2z_3)-(z_2.k_3)\epsilon(z_1k_1z_3)\\
& Y_3 \epsilon(z_1k_1k_2) = (z_3.k_1)\epsilon(z_1k_1k_2) = k^2_1\epsilon(z_1z_3k_2)+ k_1 k_2 \epsilon(z_1 k_1 z_3) = -k_1 V_2
\end{align}
to obtain
\begin{align}
Y_2Y_3[O_{\epsilon}]_1G = -Y_2(Z_1 V_2 + Y_3 W_1)\label{penstep}
\end{align}
Now we use the Schouten identities \cite{Conde:2016izb}
\begin{align}
W_1 Y_2 Y_3 +V_1(G+Y_1Z_1) = 0 \quad V_1 Y_1 = V_2 Y_2 = V_3 Y_3
\end{align}
in (\ref{penstep}) and simplify to obtain
\begin{align}
Y_2Y_3[O_{\epsilon}]_1G = G V_1.
\end{align}

\section{Net-helicity zero amplitudes and CFT correlator}\label{zeroamp}
The amplitudes in (\ref{nonpamp}) were determined by imposing little group scaling and vanishing of the amplitude in case of real momenta. This left out the case of $h_1 s_1 + h_2 s_2 + h_3 s_3 = 0$ which we consider here
\begin{align}
\mathcal{A}^{s_{1}, s_{2}, s_{3}}_{h_1, h_2, h_3}= \begin{cases}\langle 1,2\rangle^{2h_{3}s_3}\langle 3,1\rangle^{2h_{2}s_2}\langle 2,3\rangle^{2h_{1}s_1} \\ [1,2]^{-2h_{3}s_3}[3,1]^{-2h_{2}s_2}[2,3]^{-2h_{1}s_1} \label{lgfmcs}
\end{cases}
\end{align}
Notice how this case is ambiguous as it can be written in terms of either angle-brackets or square brackets. However, due to momentum conservation, we either have $[ij] = 0$ or $\langle ij\rangle = 0$, which in the above case leads to $0^0$ for one of the representations which is ill-defined. However, this is due to the fact that momentum conservation is imposed in the end, if one were to impose momentum conservation before the little group scaling, then we have
\begin{align}
\mathcal{A}^{s_{1}, s_{2}, s_{3}}_{h_1, h_2, h_3}= \begin{cases}\langle 1,2\rangle^{2h_{3}s_3}\langle 3,1\rangle^{2h_{2}s_2}\langle 2,3\rangle^{2h_{1}s_1} &\text{ when } [ij] = 0 \\ [1,2]^{-2h_{3}s_3}[3,1]^{-2h_{2}s_2}[2,3]^{-2h_{1}s_1} &\text{ when } \langle ij \rangle = 0\label{lgsmcf}
\end{cases}
\end{align}
which are well-defined. This means that the momentum conservation and little group scaling do not commute. But again, in the limit where the brackets vanish, we get a $ 0^0 $, which means that such amplitudes may not vanish for real momenta. However, such amplitudes have been shown to be inconsistent with locality and unitarity \cite{McGady:2013sga}.

Let us consider an example of net helicity zero CFT correlator.
\subsection{$\langle JJT\rangle$}
In \eqref{112msmt} it was shown that in spinor helicity variables the following CFT correlators vanishes
\begin{equation}
\langle J_{-}J_{-} T_{+}\rangle = \langle J_{+}J_{+} T_{-}\rangle =0
\end{equation}
However, corresponding flat space spinor helicity amplitude is non-zero. It is interesting to note that one can write down additional homogeneous structures for this CFT correlator and they are given by
\begin{equation}\label{112misa1}
\langle J_{-}J_{-} T_{+}\rangle = k_3 \frac{\langle 12\rangle^4}{\langle 23\rangle^2\langle 31\rangle^2}
\end{equation}
and its conjugate.
It can be verified that they satisfy
\begin{align}
K^{\kappa} \frac{ \langle J_{-}J_{-} T_{+}\rangle }{k_3} = 0 \quad K^{\kappa} \frac{ \langle J_{+}J_{-} T_{-}\rangle }{k_3} = 0
\end{align}
This in the flat space limit gives
\begin{equation}
\lim_{E\to 0}\langle J_{-}J_{-}T_{+}\rangle = k_3 \frac{\langle 12\rangle^4}{\langle 23\rangle^2\langle 31\rangle^2},~~~
\end{equation}
which precisely reproduces the extra spinor-helicity amplitudes. Converting \eqref{112misa1} in momentum space using the ansatz
\begin{align}
\langle J(z_1, k_1) J(z_3, k_3)T(z_3, k_3)\rangle &= A (z_3.k_2)^2 z_2.k_3 z_1.k_3 + B_i z_2.z_1 (z_3.k_2)^2 + C z_3.z_2 z_3.k_2 z_1.k_3 \notag\\&+ C(k_1 \leftrightarrow k_2) z_1.z_3 z_3.k_2 z_2.k_1 \label{momsp}
\end{align}
we get
\begin{align}\label{abc112}
&A = \frac{k^2_3}{8 E (E-2k_2)^2(E-2k_1)^2}
, B = -\frac{k^3_3 (E-2k_3)}{16E (E-2k_2)^2(E-2k_1)^2}, C =\frac{k^2_3k_1(E-2k_3)}{16 E (E-2k_2)^2(E-2k_1)^2}
\end{align}
In the flat space, limit (\ref{momsp}) gives
\begin{align}
\lim_{E\to 0}\langle J(z_1, k_1) J(z_2, k_2) T(z_3, k_3)\rangle &= \frac{k^2_3 z_3.k_2}{128 E k^2_2k^2_1}\bigg(-k_3 k_2 z_1.z_3 z_2.k_1+ k_1k_3 z_3.z_2 z_1.k_2\notag\\&+ z_3.k_2 z_2.z_1 k^2_3-z_3.k_2 z_2.k_1 z_1.k_2\bigg) + O(E^0)
\end{align}
which cannot be re-written as a $4D$ Lorentz invariant structure. Another issue is that this CFT structure \eqref{momsp}, \eqref{abc112} has poles of the form $k_i+k_j-k_k$ which is unphysical for a local CFT. Therefore, if one abandons locality in the CFT one can write down CFT structures that reproduces the full amplitude in the flat-space limit.

	\providecommand{\href}[2]{#2}\begingroup\raggedright
	\bibliography{refs}
	\bibliographystyle{JHEP}
	\endgroup

\end{document}